\documentclass[twocolumn]{aastex631}
\usepackage{graphicx}
\usepackage{dcolumn}
\usepackage{bm}
\usepackage{amsmath}
\usepackage{hyperref}
\usepackage{CJK}

\def\rhocnm{\rho_{\rm CNM}}
\def\Csgr{C_{\rm sgr}}
\def\mhcm{m_{\rm H}~{\rm cm}^{-3}}
\def\msun{M_{\odot}}
\def\cmc{{\rm cm}^{-3}}
\def\epe{\epsilon_{\rm e}}
\def\epb{\epsilon_{\rm B}}

\def\gmm{\gamma_m}
\def\gme{\gamma_e}

\def\Mw{\dot{M}_{\rm out}}
\def\ergs{{\rm erg~s^{-1}}}
\def\be{\begin{equation}}
\def\ee{\end{equation}}

\shorttitle{Radio Afterglows I:  Forward Shock}
\shortauthors{Mou}

\begin{document}
\bibliographystyle{apj}
\title{Numerical Studies on the Radio Afterglows in TDE:  Forward Shock}

\author{Guobin Mou (\begin{CJK}{UTF8}{gbsn}牟国斌\end{CJK}) }  
\affiliation{Department of Physics and Institute of Theoretical Physics, Nanjing Normal University, Nanjing 210023, China; gbmou@njnu.edu.cn } 
\affiliation{Nanjing key laboratory of particle physics and astrophysics, China } 


\begin{abstract} 
{ Recent long-term radio monitoring of tidal disruption events (TDEs) suggests that radio afterglows are common. Most studies argue that these afterglows may arise from forward shocks (FS) produced by the interaction between the TDE outflow and the hot, diffuse circumnuclear medium (CNM).
Current theoretical models do not model the evolution of relativistic electrons in space, which introduces uncertainties. 
Here we conducted hydrodynamic simulations to study the spatial evolution of relativistic electrons, and calculated the synchrotron spectra via radiative transfer. 
We focus on the FS scenario with non-relativistic outflows, and various parameters of the outflow and CNM are explored. 
A moderate outflow with kinetic energy of several $10^{50}$~erg in a Galactic center - like CNM can produce mJy-level radio afterglows at a distance of $100$~Mpc. The self-absorption frequency exhibits a slow decline at early times and a rapid decrease at late times. We derived the temporal evolution of the high-frequency radio flux, revealing its characteristic rise and decline pattern.  
We also find that: (1) the radio spectra for narrow outflows are clearly anisotropic along different sight lines; (2) the FS parameters inferred from radio spectra using conventional analytical formulas deviate significantly from those in simulations, in which the inferred shock radii are half of those from simulations, and the inferred energies are an order of magnitude lower.  } 
\end{abstract} 
 

\section{Introduction}
Some TDE candidates exhibit radio emission with luminosities of $10^{36-42}~\ergs$ and time lags spanning from days to years (\citealt{alexander2020}). Recent studies indicate that about half of all optically-selected TDEs show radio emission that rises on timescales of hundreds of days (\citealt{cendes2024}), suggesting that TDEs with radio emission delayed by years are prevalent. 

These radio afterglows of TDEs arise from synchrotron radiation of cosmic ray electrons (CRes), though there are several possible sources for these electrons. Some studies argue that they could come from shocks driven by relativistic jets (\citealt{bloom2011, burrows2011, zauderer2011, giannios2011, lei2016}). However, observations over the past decade show that jetted TDEs are rare \citep{andreoni2022}. 
In most current observational studies, CRes are interpreted as originating from the forward shock (FS) when wide or narrow nonrelativistic outflow interacting with hot and diffuse CNM \citep{alexander2016}. Alternatively, some studies suggest that CRes may originate from shocks produced by interaction between the unbound debris and CNM \citep{krolik2016, yalinewich2019} or dense torus \citep{lei2024}, or collisions between TDE outflows and ambient clouds \citep{mou2021b, mou2022, zhuang2025}. The former scenario essentially belongs to the FS scenario, focusing on the shock sweeping through the material ahead of the shock front. In contrast, the latter corresponds to the bow shock (BS) scenario, which focuses on the shock sweeping through the outflow. We leave the hydrodynamic simulation results for the BS scenario to a forthcoming paper \citep{mou2025b}. 

Although the origin of the radio emission is still debated, the presence of TDE outflows inevitably leads to the formation of a FS. The TDE outflow has been demonstrated to be both powerful and high-velocity. The outflow could be generated in
the self-interaction process due to GR 
precession \citep{sadowski2016, lu2020}, circularization process of infalling debris \citep{steinberg2022}, or the final accretion process \citep{dai2018,curd2019,bu2023,hu2024}. The ratio of mass outflow rate to the mass accretion rate could be considerable (up to 1/10, e.g., \citealt{steinberg2022, thomsen2022}), and accordingly, the mass of the outflow could reach $10^{-2}-10^{-1}\msun$ for disrupting a sun-like star. 
The existence of high speed outflow has been demonstrated by X-ray and UV observations of TDEs \citep{kara2018, hung2019, xiang2024}, with some blueshifted absorption lines indicating velocities of up to 0.2c. 
In this context, we first focus on investigating the FS scenario to explore the general characteristics of its radio emission. 

Theoretical investigations usually employ analytic methods (\citealt{duran2013, matsumoto2021}), or conduct hydrodynamic simulations (\citealt{hu2025}) that do not incorporate relativistic electrons in simulations. 
Due to the simplifications, the spatial distribution and evolution of relativistic electrons remains largely unknown, which introduces uncertainties in connecting shock physics with the radio spectra. The complexity of the fluid behavior therefore makes numerical simulations incorporating relativistic electrons indispensable. 

In this study, we employ a shock-injected CRe approach, utilizing hydrodynamical simulations to investigate the spatial evolution of CRe after the shock acceleration. Radiative transfer calculations are then used to calculate the radio spectra along different directions. This study reveals the relationship between the physical parameters of the FS and the resulting radio emission, and highlights the errors that 
were overlooked in previous studies. 

In Section 2, we introduce the physics related to the models. We introduce the settings of the simulations in Section 3, and present the data processing methods in Section 4. Results are presented in Section 5 and we give a brief summary in Section 6.

\section{Physics of the Models} 

\subsection{CNM}
Up to now, observational constraints on the hot CNM are available for two sources: Sgr A* and M87.
The density of CNM around Sgr A* follows $\rho(r) \simeq 30~ \mhcm r^{-1}_{-1}$ ($r_{-1} \equiv r/0.1$pc, \citealt{xu2006,gillessen2019}). 
For M87, although the BH mass is three orders of magnitude higher than Sgr A*, its CNM density at a given distance in parsec is only one order of magnitude higher than that of Sgr A* \citep{russel2015}.  
Here we simply assume a CNM density following a power-law form similar to that of Sgr A*: 
 \be
 \rho(r)=\Csgr \times 30 ~\mhcm  r^{-n}_{-1} 
 \ee
where $\Csgr$ is set to be 3.0 in the fiducial case, indicating a density 3 times that at 0.1 parsec from Sgr A*, and $n$ is the density index which is set to be 1.0 in the fiducial case. 
We also investigated different power-law indices as well (Table 1). 
Furthermore, the density profile may present different slopes inside and outside the position where the stellar wind could marginally escape the gravitation potential of the SMBH \citep{generozov2017}.  
Thus, we examined broken power-law cases, where the dividing radius between the ``inner'' and ``outer'' regions is denoted as $r_{\rm turn}$ (runs Knbr, Lnbr).  

\subsection{TDE Outflow} 

TDE is a transient phenomenon, and we assume that the outflow persists for 1 year. We constrain the outflow to a biconical structure with a half-opening angle of $\theta_0$, and inject it into the simulation domain along the two polar axes from the inner boundary of $r$. 
We parameterize the injected outflow with its velocity $v_{\rm out}$ and mass outflow rate $\Mw$ (Table 1), which remain constant during the outflow ejection epoch. In the fiducial model, the velocity is set to be 0.2 c, and the total outflow mass is $0.01 \msun$, with a corresponding energy (kinetically dominated) of $3.5\times 10^{50}$ erg. We also examined a more powerful case and two weaker cases for comparison.

\subsection{Shocks and Relativistic Electrons} 
The electron acceleration efficiency $\epe$ is assumed to be the fraction of the energy flux that can be dissipated at the shock (i.e., the change in the kinetic energy flux across the shock) channelled into the accelerated relativistic electrons in the downstream. 
In the frame of shock front, the expression of $\epe$ is 
\be
\epe=\frac{e_2 v_{d}}{\frac{1}{2}\rho_{i} v^3_s (1-C^{-2})}
\ee
where $e_2$ is the energy density of CRe in the downstream, $v_{d}$ is the downstream velocity, $\rho_{i}$ is the pre-shock density, and $C\equiv 4\mathcal{M}^2/(\mathcal{M}^2+3)$ is the compression ratio. When the Mach number $\mathcal{M} \gg 1$, we have $v_{d}=v_s/4$ and $C=4$. In this case, $\epe$ can be simplified to $0.6 e_2/e_{d}$, where $e_{d}$ is thermal pressure in the downstream. 

The acceleration efficiency $\epe$ is highly uncertain, and conventionally it should be lower than that of CRp. Simulations suggest that the acceleration efficiency of CRp of $\sim 10\%$ for high Mach number shocks (e.g., \citealt{caprioli2014}), and $\epe$ should not exceed 10\%. 
Here we adopt $\epe=0.03$ as the fiducial value. 
Note that the definition of $\epe$ here is slightly different from the one adopted in some literature where it is defined as $\epe ^{\prime}=E_e/E_s$, i.e., the total energy of relativistic electrons to the shock energy. The ratio between the two is $\epe ^{\prime}/\epe=(1-C^{-2})=0.94$ for $\mathcal{M} \gg 1$, and thus one does not need to distinguish them. 

In general, the spectral energy distribution of CRes follow a power-law form: $\frac{dn_{\rm cr}}{d\gme} =A_0 \gme^{-p}$, where $p$ is the power-law index and we set $p=2.5$ in most cases. This distribution is related to $e_2$ through: 
\be
e_2=\int^{2000}_2 \gamma m_e c^2 \frac{dn_{\rm cr}}{d\gme} d\gme ~,
\ee
where the integration upper limit (1 GeV) depends on the cooling and shock acceleration processes, but has little effect on the radio emission below 50 GHz discussed here. Once $e_2$ is specified in simulations, the coefficient $A_0$ and  the spectral energy distribution of CRe can be determined accordingly.  

\subsection{Magnetic Field}
The relativistic electrons and protons can induce various instabilities, and drive magnetic perturbations, and amplify the magnetic field (\citealt{bell2001, schure2012}). The magnetic pressure can be enhanced to a fraction of the ram pressure. 
Given the complexity of this process, we do not include the magnetic field in simulations, and simply assume that the ratio of magnetic field energy density to that of CRe in each mesh remains a constant, i.e., 
\be
\frac{B^2(t)/8\pi}{e_2(t)}=\frac{\epb}{\epe} ~. 
\ee 
We set $\epb=0.10$ in this study, which is higher than $\epe$.

\begin{figure}
\includegraphics[width=0.98 \columnwidth]{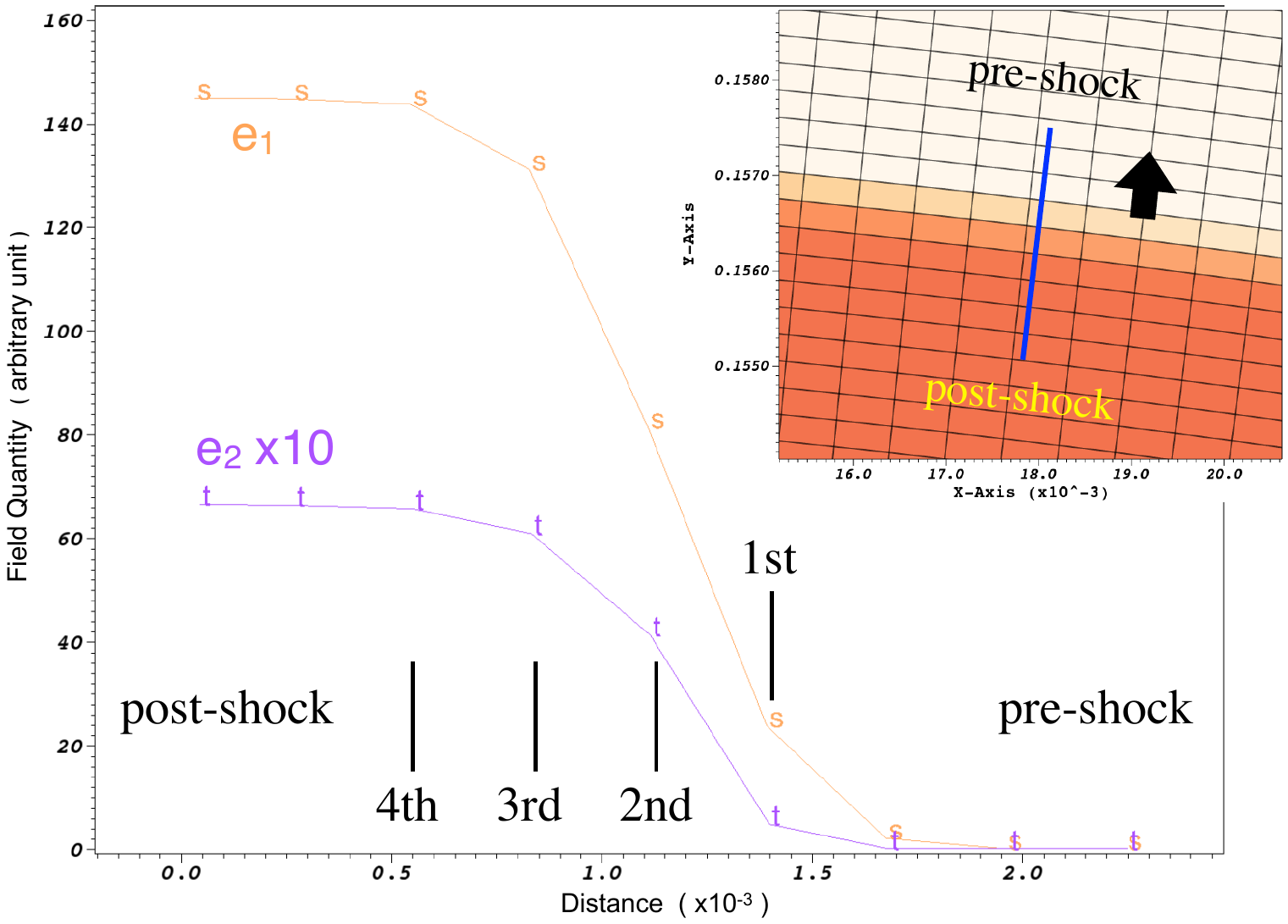}
 \caption{Schematic diagram of the CRe injection method. The energy density distributions ($e_1$ and $e_2$) refer to those in the nine meshes intersected by the blue line segment in the upper right panel. The simulation code uses 4 grid levels to capture the shock, which stabilizes at the 4th mesh (see the orange line). The CRe component is injected at the 4th mesh. Due to numerical diffusion, the values of $e_2$ in the 1st -- 3rd mesh subsequently become nonzero, but this has a negligible effect on the results.   } 
 \label{fig1}
\end{figure}

\section{Numerical settings}
We conduct the two-fluid simulation with ZEUS-3D code (\citealt{clarke2010}). 
For simplicity, we did not incorporate the magnetic field and physical diffusion of CRe. The timescale of radiative cooling of the CNM and outflow is much longer than the simulation timescale, and thus radiative cooling is negligible. The hydrodynamic equations are
\begin{gather} 
\frac{\partial \rho}{\partial t} + \nabla \cdot (\rho {\bf v})=0 , \\
\rho \frac{d {\bf v}}{d t} = -\nabla (p_{1}+p_{2}) -\rho \nabla \Phi , \\
 \frac{\partial e_{1}}{\partial t} +\nabla \cdot(e_{1}{\bf v})=-p_{1}\nabla \cdot {\bf v}, \label{hydro3} \\
  \frac{\partial e_{2}}{\partial t}+\nabla \cdot(e_{2}{\bf v}) =-p_{2}\nabla \cdot {\bf v}, 
\end{gather}
where $p_1\equiv (\gamma_1-1)e_1$ is the thermal pressure ($\gamma_1=5/3$), $p_2\equiv (\gamma_2-1)e_2$ is the pressure of CRe ($\gamma_2=4/3$), and $\Phi$ is the gravitation potential ($\Phi=-GM_{\rm bh}/r$, $M_{\rm bh}=5\times10^6~\msun$ in this work).
We did not simulate the acceleration process of electrons, but instead, directly injected the CRe component as the second fluid which is embodied as the energy density $e_2$. As mentioned in Section 2.3, $e_2$ is assigned the value of $\frac{5}{3} \epe e_{d}$, where $e_{d}$ is thermal pressure in the downstream. The shock front in ZEUS-3D is typically captured with 4 meshes, and thus we inject the CRe at the 4th-mesh (the final mesh capturing shocks) when the parameters of the post-shock gas are stabilized (figure \ref{fig1}). After that, we subtract the CRe energy density from that of the thermal gas to help maintain energy conservation. 

We adopt 2.5 dimensional spherical coordinates, in which the system is symmetric in $\phi-$direction. The biconical outflow is ejected along the two polar axes.  The inner and the outer boundary of $r$ is set to be 0.0061 pc and 0.60 pc, respectively.  Exceptions are run Bm01, Cv03, and Mm01, where the outer boundary is extended to 1.07 pc to better encompass the radio decay phase. The computation domain is divided into 2560 (or 2880 for exceptions) pieces in $r-$direction with $dr_{i+1}/dr_{i} = 1.0018$, and 256 pieces in $\theta-$ direction with $d\theta_{j+1}/d\theta_{j} = 1.004$. 
The high resolution adopted here is sufficient to ensure the results are convergent.

\begin{table*}
\centering
\caption{Parameters for modeling the radio afterglows. $\theta_0$ is the half-opening angle of the outflow. $E_{\rm k}$ is the kinetic energy of the outflow. $C_{\rm sgr}$ is the ratio of the density at $r=0.1$ pc to that at the Galactic Center. 
$F_{\rm 0.88GHz, max}$ represents the maximum value of $F_{\rm 0.88GHz}$ over time which is observed along the equatorial direction with a luminosity distance of $d_L=100$ Mpc. $F_{\rm 1.4GHz, max}$ and $F_{\rm 6GHz, max}$ are the same, but for 1.4 GHz and 6 GHz, respectively. 
For Knbr and Lnbr, we adopt two different density indices inside and outside a transition radius of 0.1 pc: $n = 1.5$ (2.5) for $r < 0.1$~pc, and $n = 0.5$ (1.0) for $r \geq 0.1$~pc. The value of $F_{\rm 6GHz, max}$ in run Lnbr represents the second flare. } 
\setlength{\tabcolsep}{0.2cm} {
\begin{tabular}{ccccccccccc}
\hline \hline
Run  & $v_{\rm out}$ & $M_{\rm out}$  & $\theta_{0}$ & $E_{\rm k}$ & $C_{\rm sgr}$  & $n$  & $F_{\rm 0.88 GHz, max}$ & $F_{\rm 1.4 GHz, max}$ & $F_{\rm 6GHz, max}$   \\
units  &  c  & $M_{\odot}$ & deg.    & erg  &  &   &  mJy & mJy & mJy \\ 
\hline
A  & 0.2  & 0.01 & 30 & $3.5\times 10^{50}$  & 3.0 & 1.0  & 3.9  & 4.0 &  1.7   \\ 
Bm10  & 0.2  & 0.10 & 30 & $3.5\times 10^{51}$  & 3.0 & 1.0  & 20  & 19 & 7.2  \\ 
Cv03  & 0.3  & 0.01 & 30 & $8.0\times 10^{50}$  & 3.0 & 1.0   & 10  & 11  & 6.1  \\ 
Dv01  & 0.1   & 0.01 & 30 & $8.9\times 10^{49}$ & 3.0 & 1.0  & 0.60 & 0.46 & 0.16 \\ 
Ev01c1  & 0.1   & 0.01 & 30 & $8.9\times 10^{49}$  & 1.0 & 1.0  & 0.17 & 0.12 & 0.04  \\ 
Fo10  & 0.2   & 0.01 & 10 & $3.5\times 10^{50}$  & 3.0 & 1.0  & 1.4 & 1.2 & 0.40  \\ 
Go60  & 0.2   & 0.01 & 60 & $3.5\times 10^{50}$ & 3.0 & 1.0  & 5.6 & 6.2 &  2.8  \\ 
Ho90  & 0.2   & 0.01 & 90 & $3.5\times 10^{50}$ & 3.0 & 1.0  & 7.3 & 8.2 &  3.4 \\ 
In15  & 0.2   & 0.01 & 30 & $3.5\times 10^{50}$ &  3.0 & 1.5 & 3.1 & 3.2 & 1.3   \\ 
Jn20  & 0.2   & 0.01 & 30 & $3.5\times 10^{50}$ & 3.0  & 2.0 & 2.6 & 2.8 & 2.8  \\ 
Knbr  & 0.2   & 0.01 & 30 & $3.5\times 10^{50}$ & 3.0  & (1.5, 0.5) & 4.7 & 5.0 & 2.4  \\ 
*Lnbr   & 0.2  & 0.10 & 90 & $3.5\times 10^{51}$ &  3.0 & (2.5, 1.0) & 45  & 49 & 25 (2nd) \\ 
*Mm01 & 0.2 & 0.10  & 30 & $3.5\times 10^{51}$ & 10.0 & 1.0 & 36 & 43 & 34  \\ 
*Nn25 & 0.2 & 0.01 & 30   & $3.5\times 10^{50}$ & 3.0   & 2.5 & 2.5 & 3.0 & 5.2  \\ 
\hline \hline  
\end{tabular} }  
{\footnotesize Runs marked with an asterisk (*) indicate that their results are presented in the Appendix.}
\label{tab:tab1} 
\end{table*}

\begin{figure*}
\centering
\includegraphics[width=1.92 \columnwidth]{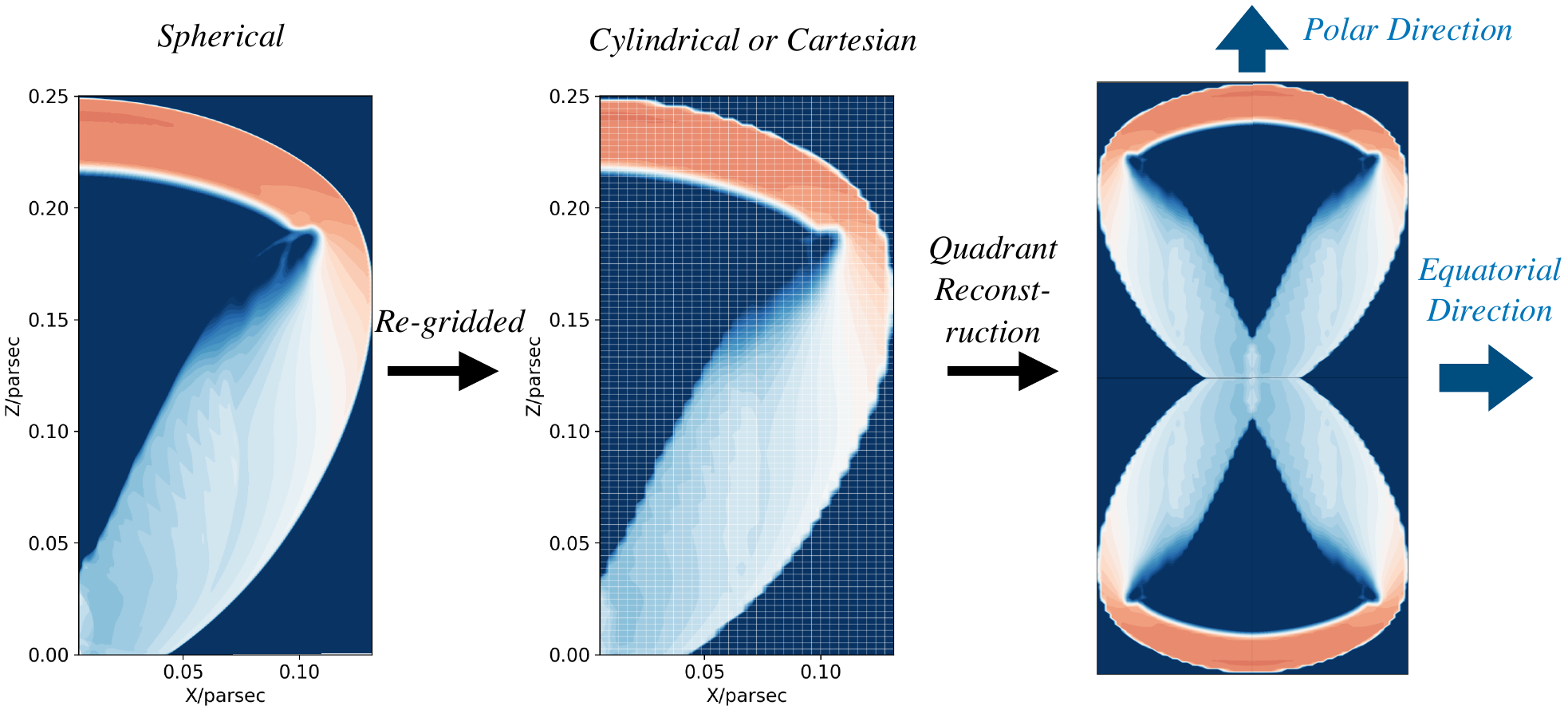}
 \caption{Schematic diagram of data processing and spectral calculation. First, the 2.5-dimensional spherical coordinates are re-gridded into a 3D coordinate system, and the CRe energy density distribution in the new coordinates is obtained via interpolation. Next, quadrant reconstruction is carried out to recover the full 3D physical scenario. Finally, synchrotron radiation transfer is performed along the polar and equatorial directions to compute the emitting spectra in both directions. } 
 \label{fig2}
\end{figure*}

\section{Calculating the radio spectra}
The simulations provide the distributions of the CRe energy density and the magnetic field strength in the study region. These parameters can be used to calculate the emitting synchrotron spectrum. 
As shown in Figure \ref{fig2}, we map the radiation zone enclosed by the FS front into a 3D Cylindrical coordinate (for polar direction spectra) and 3D Cartesian coordinate system (for equatorial-direction spectra) via interpolation. The Z-axis of the Cartesian coordinate is aligned with the polar axis and the origin is coincident with that in the spherical coordinates. 

Subsequently, we employ quadrant reconstruction to build the three-dimensional shock structure in the Cylindrical and Cartesian coordinate system. 
For 3D Cylindrical coordinates, we employ 180 meshes along Z direction, and 64 meshes along $R$-direction. For 3D Cartesian coordinates, we employed 128,  32, 72 meshes along the X, Y and Z directions, respectively. 
We have verified the convergence of the spectrum results by resolution tests.  

We solve the radiation transfer equation in the polar direction (along Z-axis, similar to ``face-on" view) and the equatorial direction (along X-axis, ``edge-on" view), respectively. 
Within one mesh, each physical parameter can be treated as uniform, and the emitting intensity can be derived from the radiation transfer equation: 
\be
\frac{\Delta I_{\nu}}{\Delta x}=-\alpha_{\nu} I_{\nu}+j_{\nu} 
\ee 
where $\alpha_{\nu}$ is the absorption coefficient, $\Delta x$ is the mesh length along the sight line, and $j_{\nu}$ is the volume emissivity.  
The solution of the intensity at the beginning of the mesh (i+1) or the end of the mesh i is: 
\be
I_{\nu}(i+1)=I_{\nu}(i) e^{-\Delta \tau_{\nu}(i)} + \frac{j_{\nu}(i)}{\alpha_{\nu}(i)}\left[1-e^{-\Delta \tau_{\nu}(i)}\right]
\ee
where $\Delta \tau_{\nu}(i)=\alpha_{\nu}(i) \Delta x_i $ is the optical depth across the $i-th$ zone. We continue the calculation iteratively to obtain the intensity from the final edge cell $x_N$ along the line of sight. 

The emitting spectra are derived by integrating the intensity from all above final edge cells of the radiation zone: 
\be
F_{\nu}=\int_k I_{\nu,N} (k) d\Omega \cos \theta
\ee
where $I_{\nu,N}(k)$ is the intensity from the $k$-th edge cell in the radiation zone, and $\theta$ is the angle between the normal direction of the radiative zone and the orientation towards the observers. 
In the polar and equatorial directions, the emitting radio spectra are calculated by: 
\begin{gather}
F_{\nu}({\rm pol})=\sum_{\substack{R}}I_{\nu,N} 2\pi R \Delta R/d^2_L   ~, \\
F_{\nu}({\rm eqt})=\sum_{Y,Z} I_{\nu,N}  \Delta Y  \Delta Z/d^2_L  ~.
\end{gather}
where $d_L$ is the luminous distance. 
In this paper, we normalize the value of $F_{\nu}$ at $d_L=100$ Mpc, corresponding to a cosmological redshift of $z=0.023$. 

In the polar direction, the separation distance between the two forward shocks moving toward and away from us results in a significant light-travel time difference (LTTD, see Appendix Figure 1A). Therefore, we incorporated this correction in calculating the polar-direction spectra. For instance, in run A, we find that the radiation from the $+Z$ FS of $t_{\rm source} = 2$ yr arrives at the observer at approximately the same time as that from the $-Z$ FS of $t_{\rm source} = 1.4$ yr. Hence, for the polar radio spectrum of $t = 2$ yr, we used simulation data of the $+Z$ shock at $t = 2$ yr and the $-Z$ shock at $t = 1.4$ yr. 
Of course, the impact of the LTTD is not substantial; for the fiducial model, its effect on the monochromatic flux is within 20\%. The LTTD is less significant in the equatorial direction; hence, we did not incorporate such a correction in this direction.

\begin{figure*}
\includegraphics[width=0.7\columnwidth]{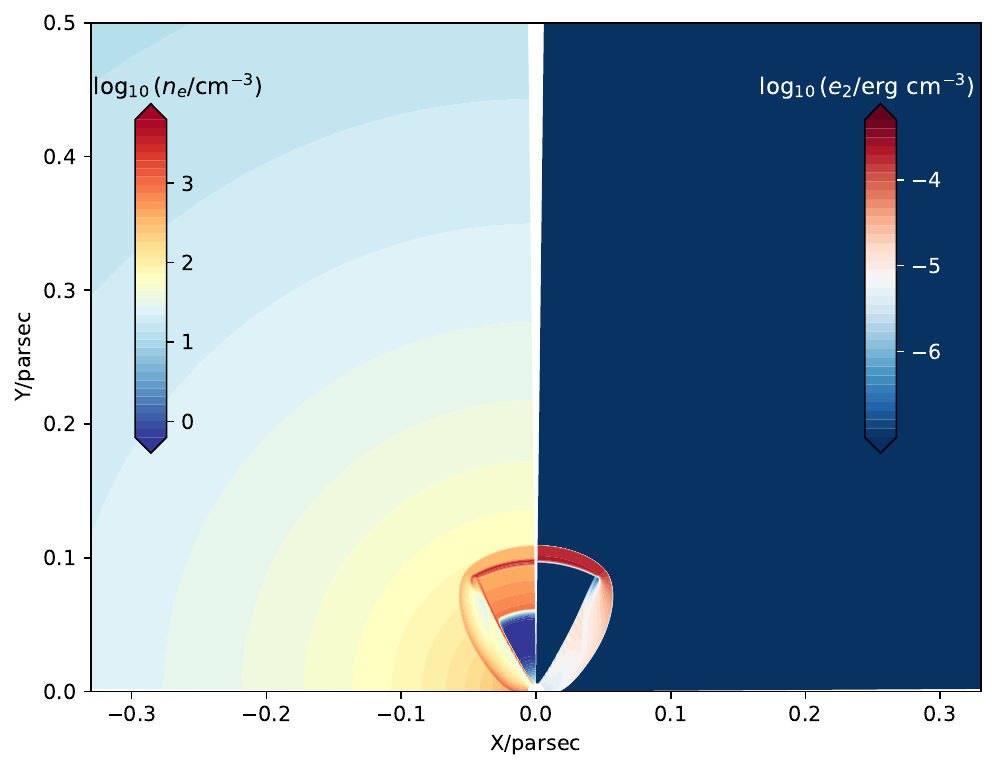}
\includegraphics[width=0.7\columnwidth]{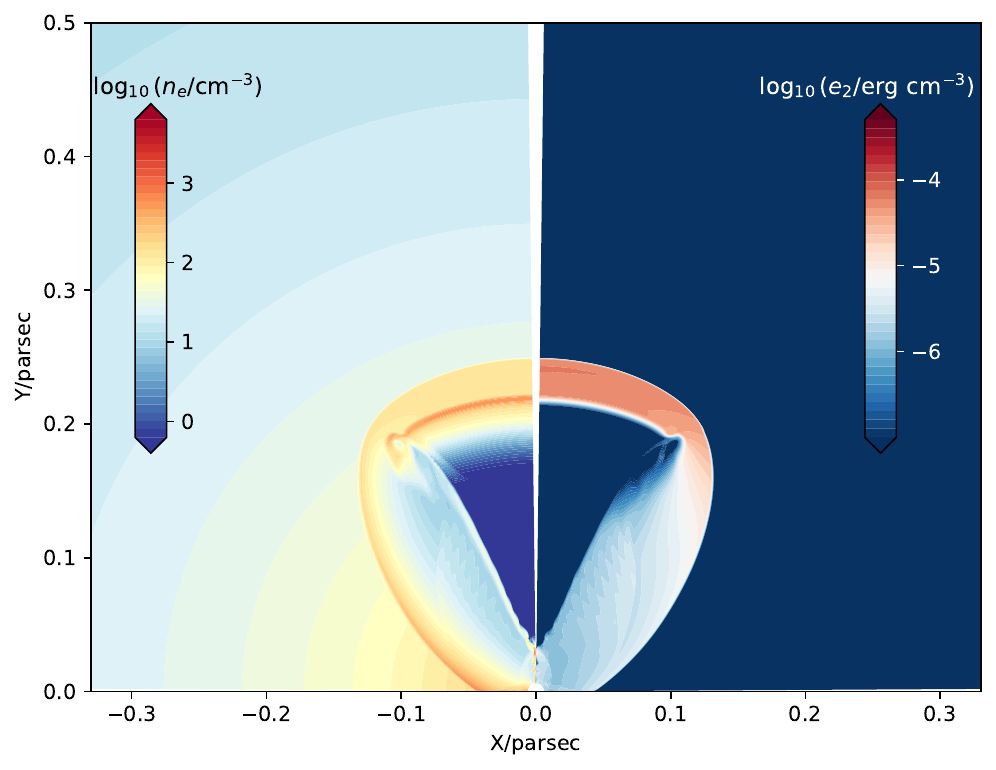}
\includegraphics[width=0.7\columnwidth]{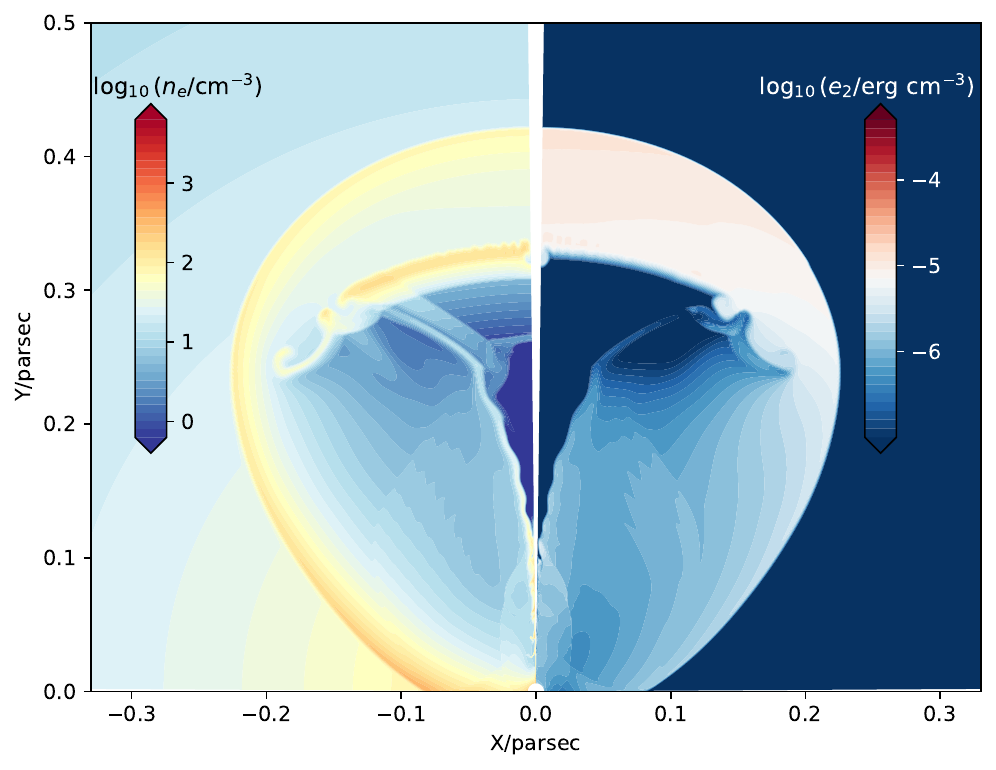}
 \caption{Snapshots of run A (fiducial run) at $t=2$ yr (\emph{left}), 5 yr (\emph{middle}) and 10 yr (\emph{right}). In each panel, the left half window shows the density distribution, and the right half window shows the energy density of CRe. } 
 \label{fig3}
\end{figure*}

\begin{figure*}
\centering
\includegraphics[width=0.79\columnwidth]{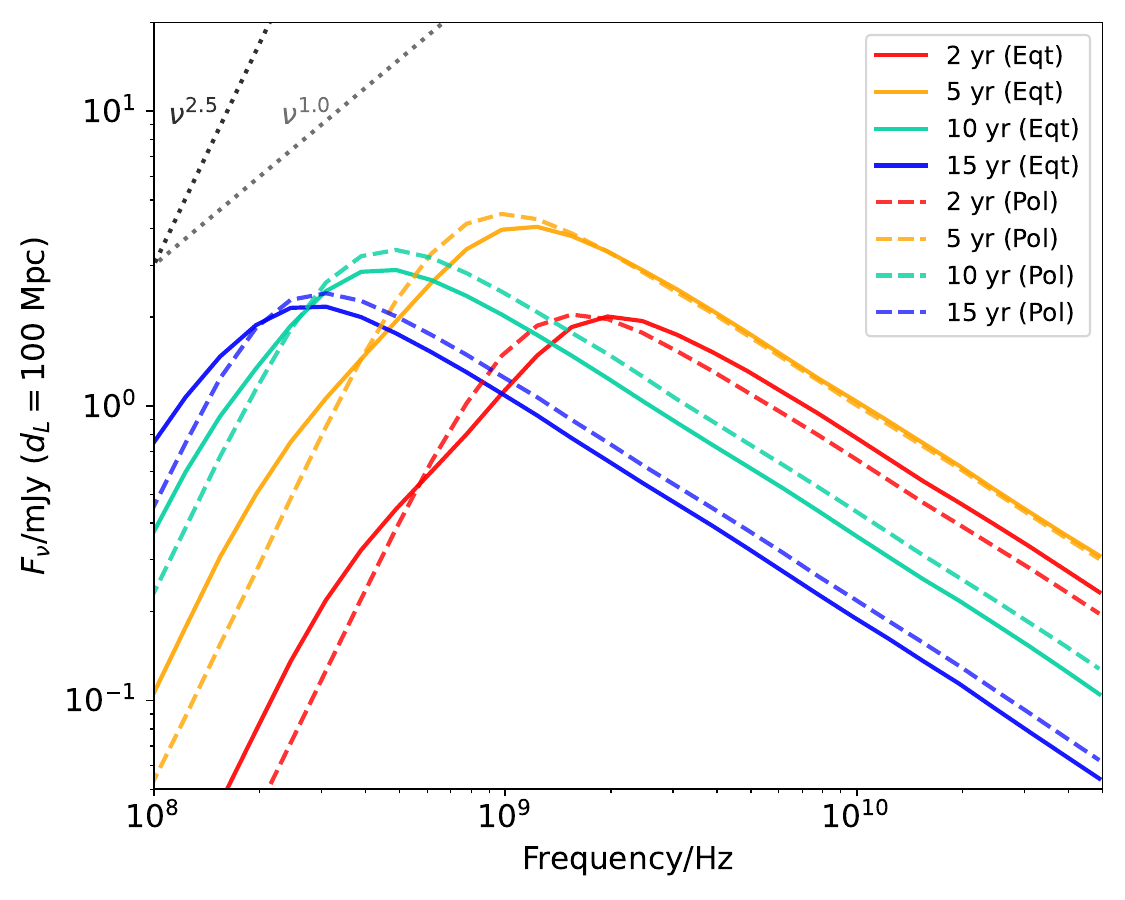}
\includegraphics[width=0.63\columnwidth]{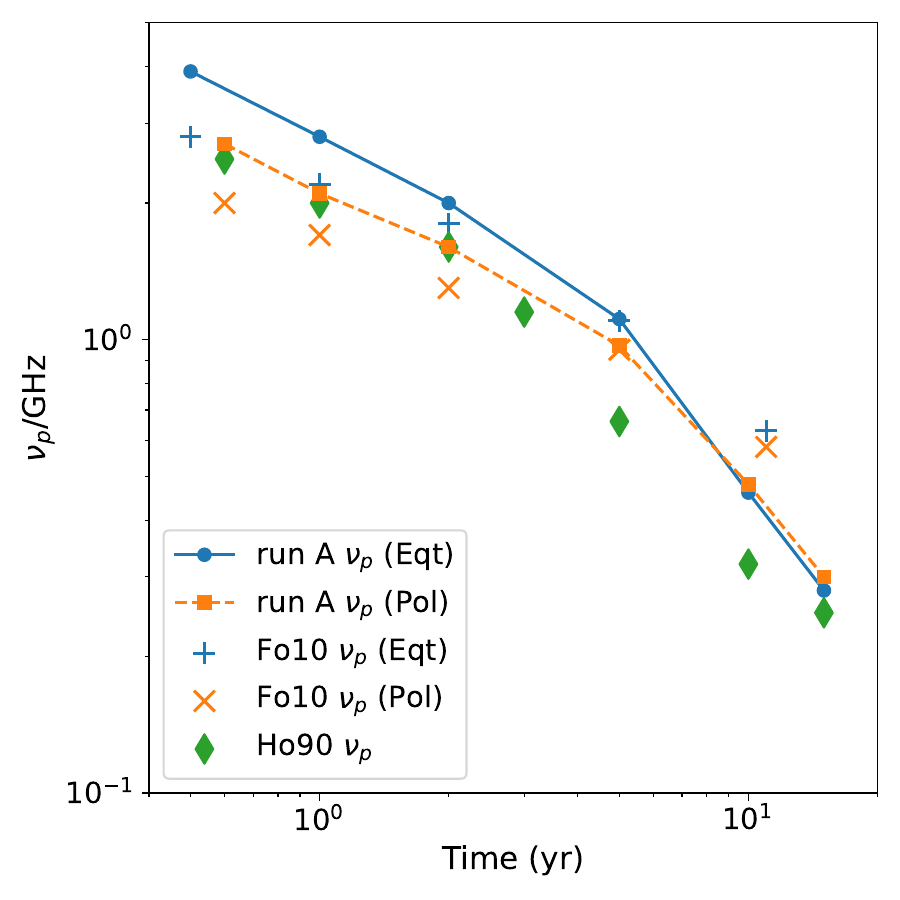}
\includegraphics[width=0.63\columnwidth]{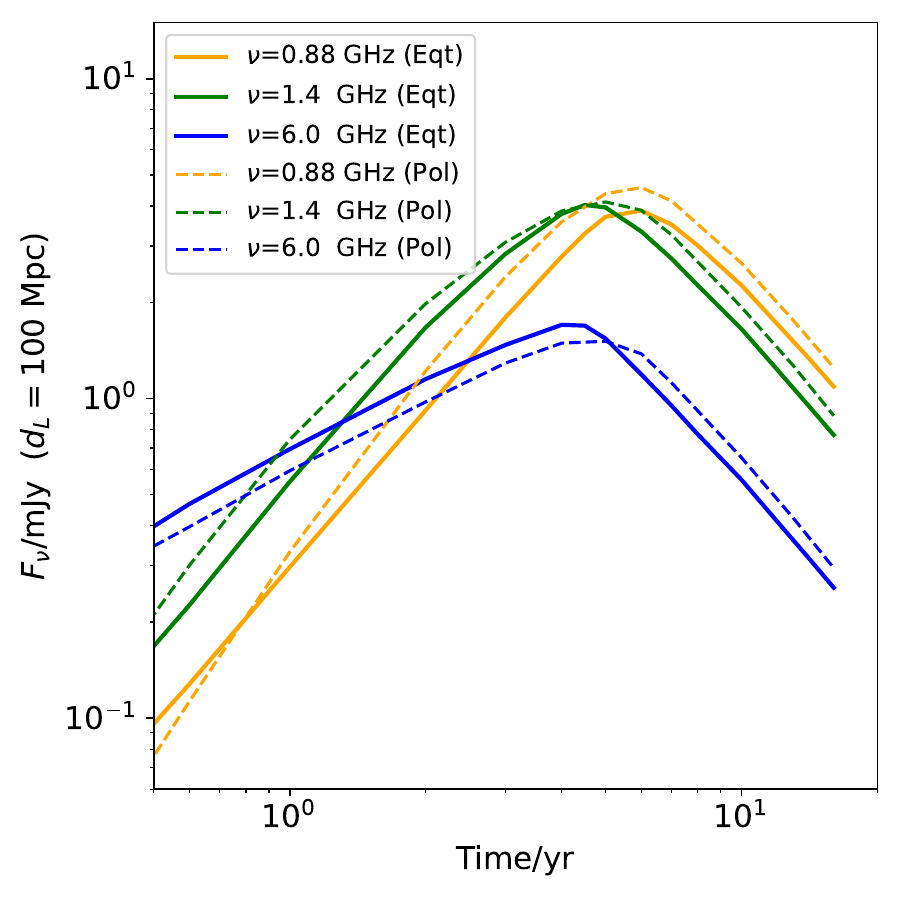}
 \caption{The left panel shows the synthetic radio spectra along the equatorial direction (Eqt) and polar direction (Pol). Note that we have considered the light-travel time difference in the polar direction. The middle panel presents the time evolution of the self-absorption frequency $\nu_p$.  
The right panel presents the light curves of $F_{\nu}$ for three frequencies. }  
 \label{fig4}
\end{figure*}

\section{Results and Discussions}

\subsection{Basic feature of the absorption frequency and monochromatic luminosity}

In the fiducial case (run A), we find that the FS scenario can produce a self-absorption frequency $\nu_p$ around GHz, with a peak monochromatic flux of several mJy. This corresponds to a peak luminosity (defined as $\nu_p F_{\nu_p}$) of $10^{37}~\ergs$. Such a luminosity is consistent with that observed in some radio TDEs (\citealt{cendes2024}). 

The self-absorption frequency of run A decreases over time (Figure \ref{fig4}), with $\nu_p(t)$ roughly following a $t^{-0.8}$ trend overall. However, the decline is more gradual at early times ($t^{-0.5}$) and becomes steeper at later times ($t^{-1.25}$). 
This two-stage evolutionary feature is also observed in other models. The turnover point corresponds to the moment when the swept CNM mass becomes comparable to the ejecta mass. However, the discrepancies in temporal indices among different models reflect the complexity of the underlying hydrodynamic behavior. 

The radio flux is affected by the FS parameters, which are determined by both the outflow and the CNM. 
When the ejecta mass is increased by a factor of 10 (run Bm01), the monochromatic fluxes at three frequencies (0.88, 1.4 and 6 GHz) rise to 4--5 times those in the fiducial case. Correspondingly, the maximum value of the peak luminosity increases to $3\times 10^{38} ~\ergs$. 
The highest flux among the single density index models is from model Mm01, in which the CNM density is 10 times that of the Galactic Center. The maximal flux reaches several tens of mJy (Table \ref{tab:tab1}). 

Compared to the fiducial model, increasing the outflow velocity (Cv03), enlarging the outflow opening angle (Go60, Ho90), or raising the CNM density all lead to significantly higher radio fluxes.  
However, our simulations indicate that cases where the monochromatic flux exceeds 10 mJy are uncommon. Generally, when the flux exceeds 10 mJy $(d_L/100 {\rm Mpc})^{-2}$, or the peak luminosity $\nu_p L_{\nu_p}$ exceeds $10^{38}~\ergs$, it may indicate that the outflow is quite powerful (exceeding $\sim 10^{51}$ erg), or that the CNM density is relatively high (significantly above that in the Galactic center). 
Moreover, none of these simulations produce high luminosities of $10^{39} ~\ergs$ or large fluxes of 100 mJy $(d_L/100 {\rm Mpc})^{-2}$, which have been detected in certain candidates such as ASASSN-15oi \citep{horesh2021} and AT2018hyz \citep{cendes2022}. Within the FS scenario, achieving such a high luminosity would require outflow's energy or CNM density significantly higher than those assumed in our simulations. Alternatively, other mechanisms beyond the FS scenario may be responsible. 
For moderate conditions, however, the maximal peak luminosity ($\nu_p L_{\nu_p}$) drops below $10^{37} ~\ergs$ (e.g., run Ev01c1 with a velocity of 0.1c and a Milky Way-like density).  

\begin{figure*}
\centering
\includegraphics[width=0.9\columnwidth]{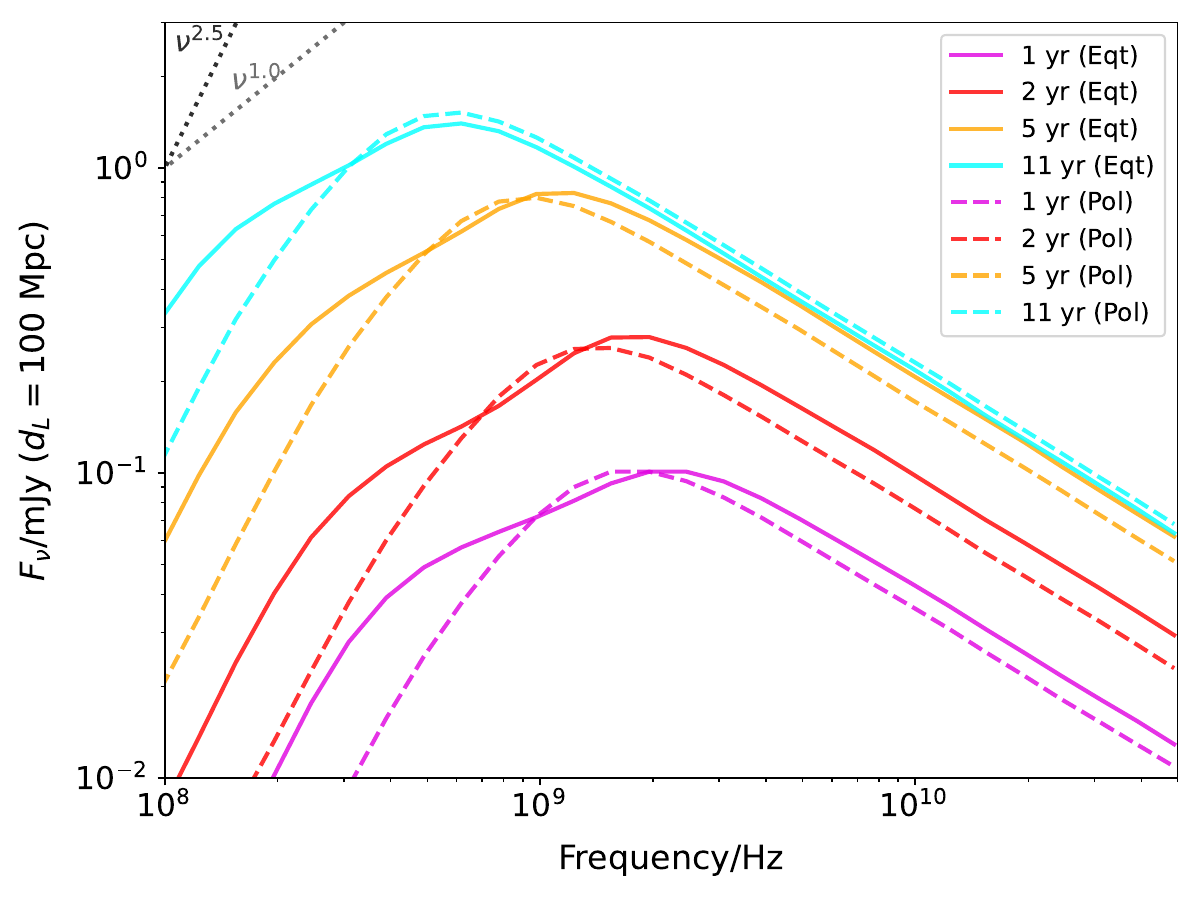}
\includegraphics[width=0.9\columnwidth]{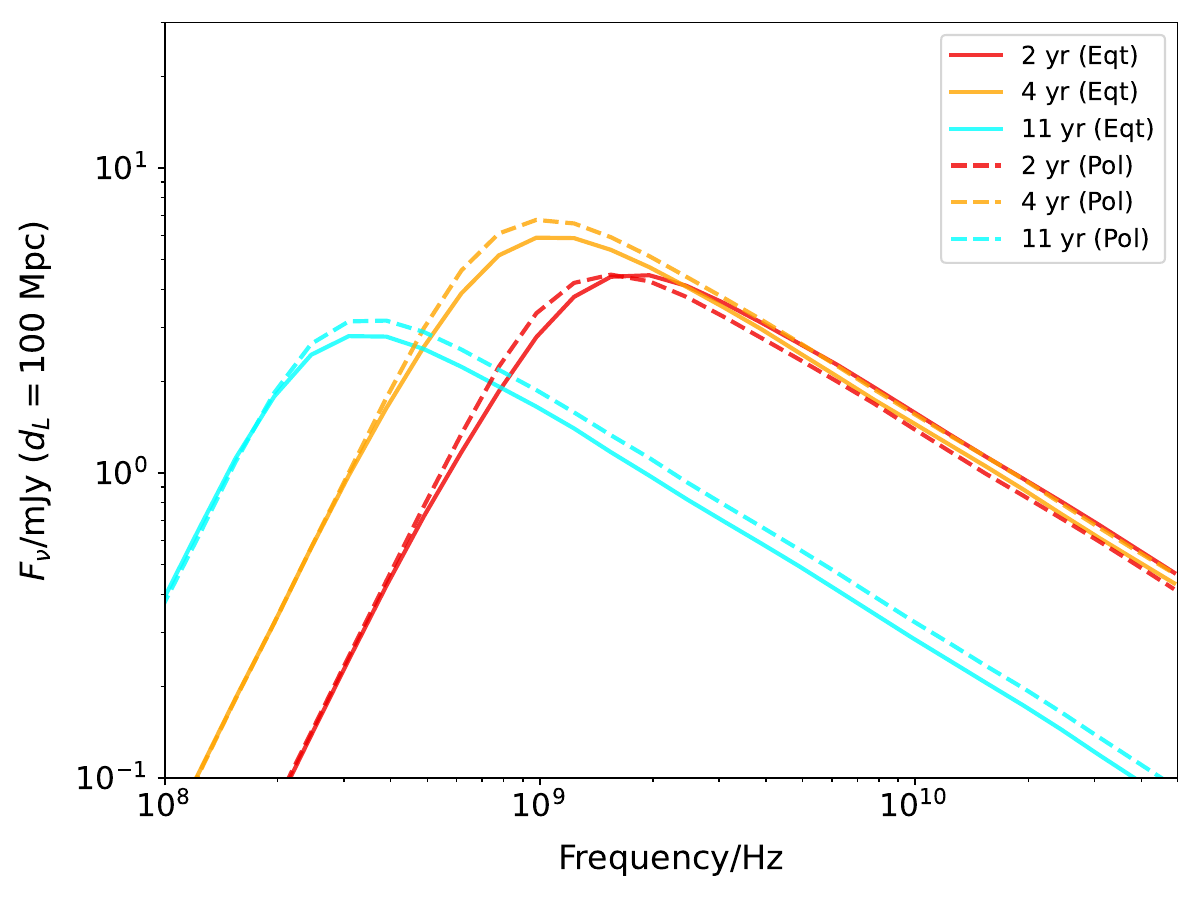}
 \caption{The emitting radio spectra for different outflow opening angles $\theta_0$. The left panel shows the spectra at different epochs for run Fo10 ($\theta_0=10^{\circ}$), and the right panel is for run Go60 ($\theta_0=60^{\circ}$). Obviously when the outflow's opening angle is smaller, the anisotropy of the spectra is more significant.  } 
 \label{fig5}
\end{figure*}

\subsection{Properties of the radio spectra} 

The spectra are predominantly contributed by the CRes at ``cap'' of the FS (the region within the outflow's opening angle), while the lateral FS also hosts a substantial population of CRes (Figure \ref{fig3}, \ref{figA2}). Such a distribution of CRes leads to anisotropy of the radio spectra (Figure \ref{fig4}). Along the equatorial direction, CRes in the lateral region of the FS contribute to low-frequency emission, leading to a broadening of the spectrum at $\nu<\nu_p$. Meanwhile, the longer path of synchrotron emission through ``cap'' region increases the optical depth,
resulting in a self-absorption frequency $\nu_p$ that is higher than in the polar direction.

The anisotropy becomes increasingly significant as the outflow opening angle decreases (Figure \ref{fig5}). In particular, when $\theta_0$ is small $\theta_0 \lesssim 10^{\circ}$, the spectra in the equatorial direction exhibit a ``flat--top'' feature -- that is, below the self-absorption frequency, the flux decreases slowly as the frequency decreases, largely deviating from the $\nu^{5/2}$ feature. 
When $\theta_0$ is large, the ``flat--top'' feature disappears, and the emitting spectra along the two directions become nearly identical. 
Therefore, multi-frequency observations could help to constrain the viewing angle and outflow opening angle.

\begin{figure}
\centering
\includegraphics[width=0.95\columnwidth]{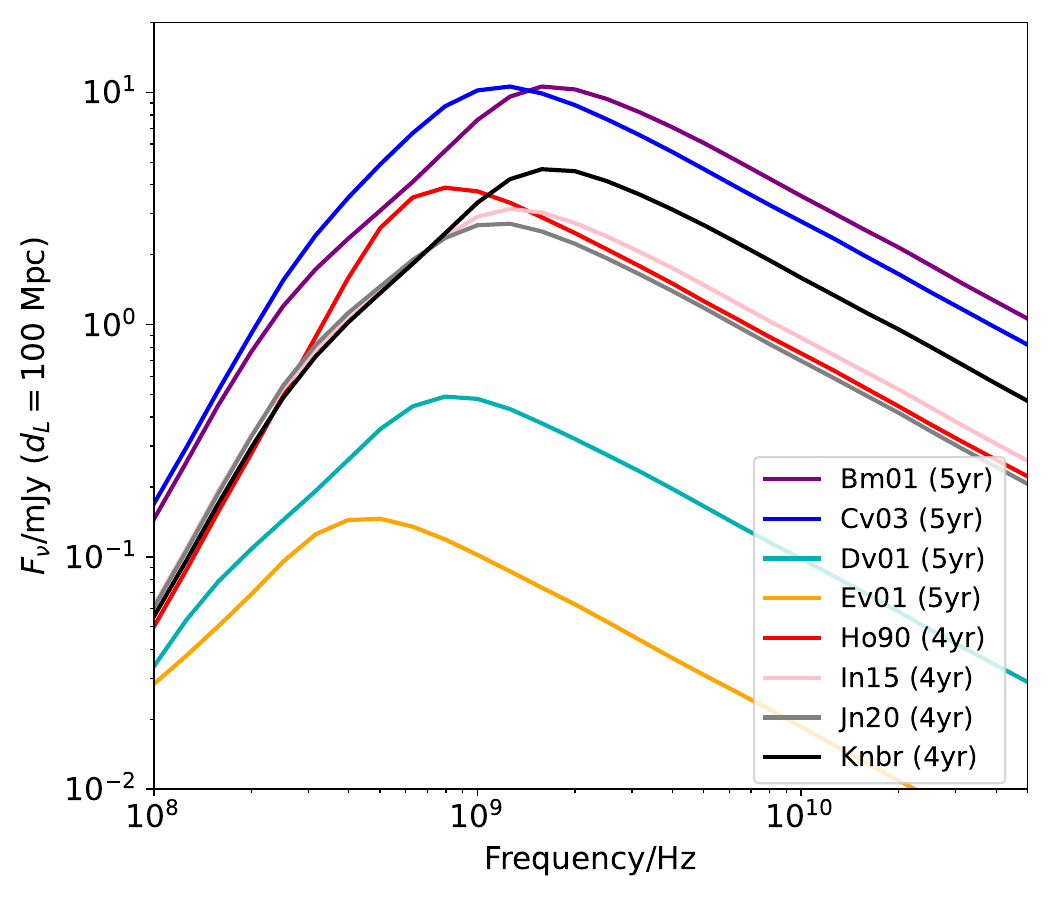} 
 \caption{The synthetic spectra for different runs, observed along the equatorial direction.  } 
 \label{fig6}
\end{figure}

Simulations provide detailed data that can be used to assess the deviations of the widely adopted energy equipartition method \citep{duran2013}.
We extract the values of $\nu_p$ and $F_{\nu_p}$ from the synthetic spectra (e.g., Figure \ref{fig6}), and then apply the equipartition method (see Appendix \ref{app2} for more details) to calculate parameters including the forward shock radius $R_{\rm eq}$, velocity $v'_s$, total number of thermal electrons $N'_e$, the density of the post-shocked CNM $n'_e$, and various energy components including relativistic electron's energy $E'_e$, the magnetic field energy $E'_B$, and the shock energy $E'_s$. Our results are listed in Table \ref{tab:A1}. 
It should be noted that $N'_e$ refers to the number of thermal electrons of the Maxwellian distribution, rather than the number of relativistic electrons accounting for the radio emission. Many observational studies mistakenly treat the number or number density of relativistic electrons as that of thermal electrons (as pointed out in \citealt{matsumoto2021}). Our $N'_e$ is derived following the sequence: $E'_e \rightarrow E'_s  \rightarrow$ proton/electron mass $ \rightarrow N'_e$ (see Appendix \ref{app2}). 

For comparison, the corresponding hydrodynamic data are presented in Table \ref{tab:tab2}. The shock energy here is the sum of the kinetic and internal energies of the post-shock CNM. We calculate the shock energy both over the entire domain marked as $E_s$, and within the outflow's opening angle marked as $E_s$(cone). 
It should be noted that at late stages, as shown in the $t=10$ yr panel of Figure \ref{fig2}, the CRe has clearly extended beyond the outflow cone. Moreover, the CNM swept up by the forward shock has undergone significant expansion, resulting in a thickness-to-radius ratio that substantially exceeds its early-time value and is also significantly greater than the commonly adopted value of 0.1. Therefore, applying the equipartition method at late times introduces additional uncertainty, and here we mainly focus on the data for $t \leq 5$ yr in most cases. 

First, the analytical estimates of the FS radius are significantly lower than those in simulations. Specifically, based on the polar direction spectra, the estimated radii are about half of those in simulation; for the equatorial direction spectra, the estimated radii are approximately $1/3 \sim 1/2$ of those in simulations (the column $R_{\rm eq}/R_s$ in Table \ref{tab:tab2}).  
In simulations, the shock radius $R_s$ is defined as distance of the shock front at the polar axis to the SMBH. 
This is motivated by the fact that radio emission is dominated by the shock region within the outflow opening angle, and within this cone, the local shock radius varies slightly from the polar value. 

Second, the estimated shock energies $E'_s$ are on average, 9 times lower than those from simulations (see the column $E'_s/E_s$ in Table \ref{tab:tab2}). 
The estimated total energy of CRe are on average 9-10 times lower than those from simulations. 
The energies derived from the analytical method are systematically underestimated, primarily due to the fact that derivation in the equipartition method involves multiple approximations, not the specified values of the $\epe$ or $\epb$ parameter.

Third, the analytically estimated densities of the post-shock CNM ($n'_e$) are higher by a factor of a few to dozens compared to those in simulations (the column $n'_e/n_e$ in Table \ref{tab:tab2}). Note that, since the derived shock radius corresponding to $n'_e$ is about half of the simulated value, the difference $n'_e$ and $n_e$ is expected to be further reduced when compared at the same radius. However, a discrepancy of several times (averaged a factor of 4-5) still remains. 

Finally, we also investigate the influence of different power law indices of CRe. We perform calculations for both $p=2.2$ and $p=2.8$, and find that the above conclusions still hold (Table \ref{tab:tab2}). 

In summary, we recommend that when estimating shock parameters using the equipartition method with specified $\epe$ and $\epb$, the derived $R_{\rm eq}$ should be multiplied by 2, the $E_e$ value by $9\sim10$, the estimated shock energy ($E_e/\epe$) by 9,  and the CNM density by $4\sim5$. Since the magnetic field energy should be multiplied by a factor of $9\sim10$ while the radius doubles, magnetic field strength remains essentially unchanged. 

The derivation of the equipartition method adopts several approximations \citep{duran2013}, the cumulative effect of which could exert a significant impact on the final results. One of the main contributing factors is the total relativistic electron energy $E_e = N_e \gamma_m m_e c^2$, missing a factor of $(p-1)/(p-2)$ which is 3 for $p = 2.5$. We speculate that the geometric factors $f_A$ and $f_V$, and the associated projected area $A$ and emitting volume $V$, are not the primary sources of the discrepancy between simulations and equipartition methods. Runs Go60 and Ho90, which have large opening angles and shock geometries closer to spherical, yield discrepancies consistent with those of the narrower-angle runs (Table \ref{tab:tab2}), confirming that the underestimation of shock radii and energies by equipartition methods is not driven by geometric assumptions. 

\subsection{Properties of light curves}

First, we considered the early stage in which the mass of the post-shock CNM is much lower than the ejecta mass, and the forward shock velocity $v_s$ can be treated as a constant.  
The mass swept-up rate of the CNM by the FS is $ \dot{M}_{\rm CNM}= \rhocnm (r) v_s ~\Omega r^2 $
where $\Omega$ is the solid angle of the outflow. 
The total shock energy is 
\be
E_{\rm s}(t) = \int^{t}_{0} \frac{1}{2} \dot{M}_{\rm CNM} v^2_{\rm s} dt = \frac{1}{2(3-n)} \rho_0 \Omega v^{5-n}_s t^{3-n} ~.
\ee  
The total energy of CRe is $E_e(t)\simeq \epe E_{\rm s}(t) \propto t^{3-n}$. 
On the other hand, the magnetic field strength can be estimated by 
\be
B=(8\pi \epb \rho_{\rm CNM} v^2_s)^{1/2}=(8\pi \epsilon_{\rm B} \rho_0 v^{2-n}_s t^{-n})^{1/2} \propto t^{-n/2} ~.
\ee 
The typical frequency of synchrotron radiation from an electron with the Lorentz factor $\gamma_m$ is 
\be
\nu_m = \frac{\gamma^2_m eB}{2\pi m_e c} \propto B \propto t^{-n/2} ~.
\ee
The spectral power at frequency $\nu_m$ from those electrons with the Lorentz factor $\gamma_m$ is 
\be
L_{\nu_m} \simeq  \frac{N_e}{\nu_m} \frac{4}{3} \sigma_T c \gmm^2 \frac{B^2}{8\pi} \propto t^{3-1.5n}~. 
\ee
Thus, for a fixed frequency $\nu_i > \nu_{p}$, the monochromatic luminosity is
\be
L_{\nu_i}=L_{\nu_m} \left( \frac{\nu_i}{\nu_m} \right)^\frac{1-p}{2} \propto t^{~\Gamma_1}
\ee
where the time index $\Gamma_1 \equiv 3-0.25np-1.25 n$.

\begin{figure}
\includegraphics[width=0.98\columnwidth]{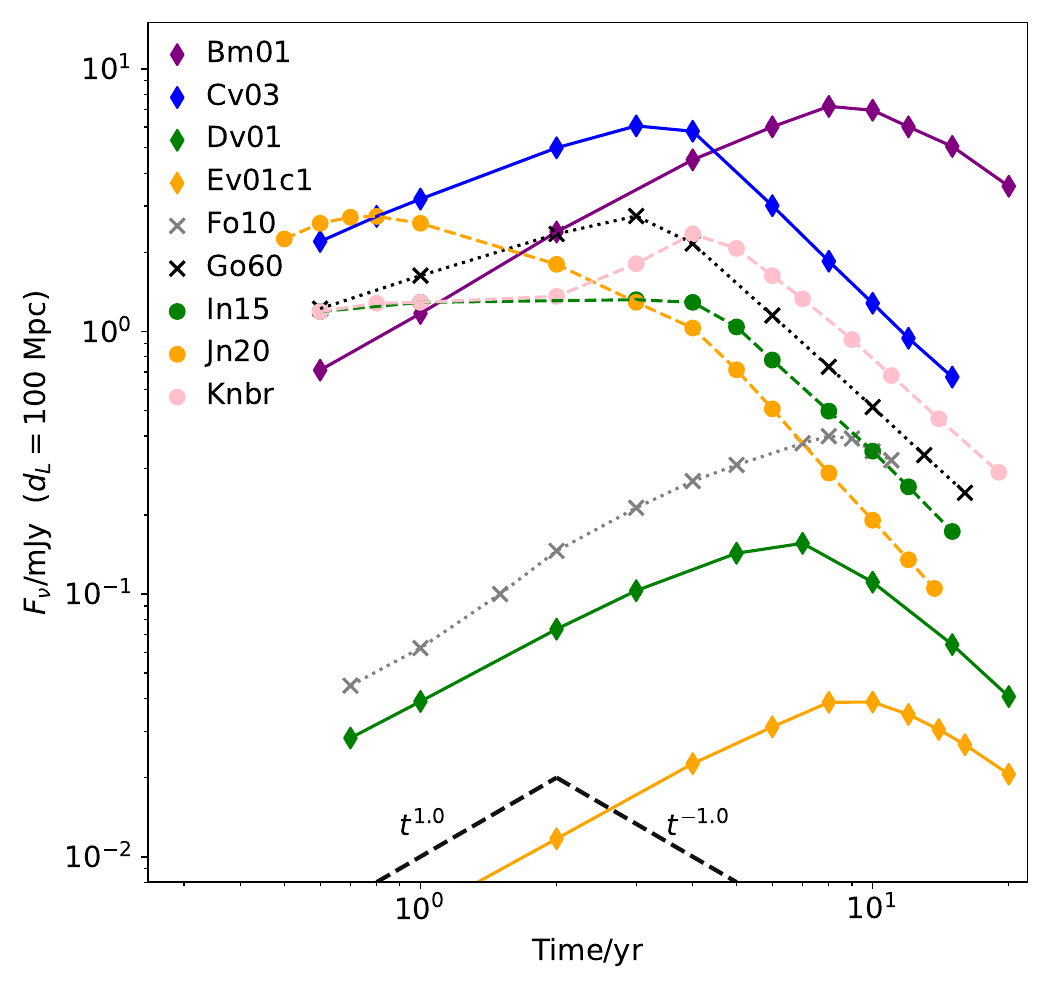}
 \caption{Temporal evolution of the monochromatic flux (6 GHz) in the equatorial direction. In most cases, the flux exhibit a rising-then-falling trend. The maximum flux occurs when the mass of the post-shock CNM becomes comparable to the outflow mass, specifically in the range of 0.4 to 1.7 times the ejecta mass based on our different simulation runs. }  
\label{fig7}
\end{figure}

We compared this analytical result with the simulation results (Figure \ref{fig4} and \ref{fig7}) and found that they are generally in agreement. 
Based on these results, for most parameter values ($p > 2$, $n \geq 1$), we have $\Gamma_1 \lesssim 1.25$, suggesting that the high frequency radio flux generated by the FS exhibits a more gradual evolution over time than $t^{1.25}$. 
Interestingly, for density index of $n > 3/(0.25p+1.25)$, $L_{\nu_i}$ is expected to decrease monotonically with time -- the radio flux would only become fainter and fainter after the very early epoch (Figure \ref{fig7}, run Jn20). 

Second, we consider the late stage in which the post-shock CNM's mass significantly exceeds the outflow mass. In this stage, the kinetic energy of post-shock CNM approaches saturation: $\frac{1}{2} M_{\rm cnm}(r) \left[\frac{3v_s(r)}{4} \right ]^2 \simeq$ constant, where $M_{\rm cnm}$ is the swept mass by the FS. The velocity variation and the newly swept-up CNM mass satisfy the relationship: $dv_s(r)/v_s(r)=-\frac{1}{2}dM_{\rm cnm}(r)/M_{\rm cnm}(r)$. From this equation, we derived that $v_s \propto r_s^{\frac{n-3}{2}}$, $r_s \propto t^{\frac{2}{5-n}}$, and $v_s \propto t^{\frac{n-3}{5-n}}$. 
The total shock energy in this stage can be roughly regarded as a constant, and the same is true for the total energy of CRe. On the other hand, since the magnetic pressure is proportional to the shock ram pressure, the magnetic field strength scales as $B\propto t^{-\frac{3}{5-n}}$. Thus in the late stage, $L_{\nu_i}$ scales as 
\be
L_{\nu_i} \propto A_0(t) B^{\frac{p+1}{2}}(t) \propto t^{~\Gamma_2}
\ee
where the index $\Gamma_2 \equiv -\frac{3p+3}{2(5-n)}$.   

We compared this analytical result with the simulation results and found that they are generally in agreement. Specifically, values of $\Gamma_2$ from simulations are slightly lower (in terms of absolute magnitude, slightly larger), meaning that real decline of flux is slightly faster than the above expression. 

Such a temporal evolving law of the high frequency ($\nu > \nu_p$) radio flux provides a way for constraining the CNM density index $n$.  

In certain cases with particular density distributions, a second flare in high-frequency flux may emerge. 
For example, in run Lnbr ($n=2.5$ for $r<0.1$ pc and $n=1.0$ for $r\geq 0.1$ pc) which is similar to the case explored in \citet{matsumoto2024}, as the shock enters the outer region with a more gradually declining density ($t\simeq 3$ yr), the radio emission begins to increase, exhibiting a rise timescale of one year. However, the second flare is not obvious, which is only enhanced by 10\% (Figure \ref{figA4}). 
Another example is run Knbr ($n=1.5$ for $r<0.1$ pc and $n=0.5$ for $r\geq 0.1$ pc), exhibiting a rise timescale of two years and a flux enhancement of 73\%. This suggests that if the second flare is more prominent, the density index $n$ in the outer region should be lower $n \lesssim 0.5$ under the FS scenario. 
Similar phenomena -- a second brightening at high frequencies ($\nu > \nu_p$) -- have been reported in some sources, such as AT2019azh (15.5 GHz, \citealt{goodwin2022}) and AT2020vwl (5.5 GHz, \citealt{goodwin2024}). 
However, in these observations, the rise and decay timescales of the second brightening are short, with the flux halving timescale of 0.6 year, which differs from our simulations. 
Therefore, the light curves obtained from current simulations indicate that the FS light curves generally exhibit slow variability. Rapid and large--amplitude variations (by at least a factor of two) are likely caused by other mechanisms.

In Figure \ref{figA3} and \ref{figA4} of the Appendix, we present the temporal evolution of the flux at two additional single frequencies: 0.88 GHz and 1.4 GHz. These two frequencies are also commonly used in observations. For these low-frequency flux, we do not observe any second flare in any of the models. 

Finally, we also present the specific values of the shock energy and outflow energy in Table \ref{tab:tab2}. Based on simulations, we find that when the high-frequency radio flux begins to decline, the shock energy starts to saturate, and its value at that turnover moment is approximately half of the outflow energy. This empirical relation can be used to estimate the kinetic energy of the outflow. If long-term monitoring can capture the rising -- decline phase of the flux at $\nu > \nu_p$, one can infer that the outflow's kinetic energy is about $2\epe^{-1}$ times the value of $E_e$, which corresponds to $20E_e$ if $\epe = 0.1$. 

\begin{table*}[htbp]
\centering
\scriptsize
\caption{Physical parameters from simulations, and the ratios of the values obtained using the equipartition method (Table \ref{tab:A1}) to those from the simulation data. $V_s$ the instantaneous shock velocity, and $N_e$ is the total amount of electrons in the post-shock CNM within the opening angle of the outflow. In the ``Direc'' column, the labels $p2.2$ and $p2.8$ indicate that spectral indices of $p = 2.2$ and 2.8 for the CRe are explored. } 
\begin{tabular}{cccccccccccccccc}
\hline \hline
Run & $\Delta T$ & Direc & $\nu_p$  & $F_{\nu_p}$ & $R_s$  & $V_s$ & $N_e$ & $n_e$ & $E_e$ & $E_s({\rm cones})$ & $E_s$ & $R_{\rm eq}/R_s$ & $n'_e/n_e$ & $E'_s/E_s$ \\
units   & yr & --  & GHz & mJy & pc & c  &   & $\cmc$  & erg  & erg  & erg  &   &  & \\
\hline
A & 2 & Pol & 1.6 & 2.0 & 0.109 & 0.163 & $1.6\times10^{54}$ & 306 & $1.2\times10^{48}$ & $3.9\times10^{49}$ & $4.9\times10^{49}$ & 0.45 & 6 & 0.11 \\
A & 2 & Eqt & 2.0 & 2.0 & 0.109 & 0.163 & $1.6\times10^{54}$ & 306& $1.2\times10^{48}$ & $3.9\times10^{49}$ & $4.9\times10^{49}$ & 0.36 & 14. & 0.084 \\
A & 2 & Eqt $p2.2$ & 2.3 & 3.1 & 0.109 & 0.163 & $1.6\times10^{54}$ & 306 & $1.2\times10^{48}$ & $3.9\times10^{49}$ & $4.9\times10^{49}$ & 0.39 & 9 & 0.067 \\
A & 2 & Eqt $p2.8$ & 1.7 & 1.1 & 0.109 & 0.163 & $1.6\times10^{54}$ & 306 & $1.2\times10^{48}$ & $3.9\times10^{49}$ & $4.9\times10^{49}$ & 0.32 & 22 & 0.086 \\
A & 5 & Pol & 0.98 & 4.4 & 0.25 & 0.144 & $3.3\times10^{54}$ & 88 & $5.1\times10^{48}$ & $1.4\times10^{50}$ & $1.9\times10^{50}$ & 0.46 & 7 & 0.12 \\
A & 5 & Eqt & 1.11 & 4.1 & 0.25 & 0.144 & $3.3\times10^{54}$ & 88 & $5.1\times10^{48}$ & $1.4\times10^{50}$ & $1.9\times10^{50}$ & 0.40 & 12 & 0.095 \\
A & 5 & Eqt $p2.2$ & 1.3 & 6.2 & 0.25 & 0.144 & $3.3\times10^{54}$ & 88 & $5.1\times10^{48}$ & $1.4\times10^{50}$ & $1.9\times10^{50}$ & 0.41 & 9 & 0.068 \\
A & 5 & Eqt $p2.8$ & 1.0 & 2.3 & 0.25 & 0.144 & $3.3\times10^{54}$ & 88 & $5.1\times10^{48}$ & $1.4\times10^{50}$ & $1.9\times10^{50}$ & 0.34 & 24 & 0.089 \\
A & 10 & Pol & 0.48 & 3.4 & 0.422 & 0.087 & $2.5\times10^{55}$ & 30 & $8.3\times10^{48}$ & $1.4\times10^{50}$ & $2.2\times10^{50}$ & 0.50 & 6 & 0.15 \\
A & 10 & Eqt & 0.46 & 2.9 & 0.422 & 0.087 & $2.5\times10^{55}$ & 30 & $8.3\times10^{48}$ & $1.4\times10^{50}$ & $2.2\times10^{50}$ & 0.48 & 7 & 0.13 \\
A & 10 & Eqt $p2.2$ & 0.52 & 4.5 & 0.422 & 0.087 & $2.5\times10^{55}$ & 30 & $8.3\times10^{48}$ & $1.4\times10^{50}$ & $2.2\times10^{50}$ & 0.52 & 4 & 0.10 \\
A & 10 & Eqt $p2.8$ & 0.4 & 1.7 & 0.422 & 0.087 & $2.5\times10^{55}$ & 30 & $8.3\times10^{48}$ & $1.4\times10^{50}$ & $2.2\times10^{50}$ & 0.43 & 10 & 0.14 \\
Bm01 & 5 & Pol & 1.25 & 10.7 & 0.301 & 0.186 & $1.1\times10^{55}$ & 110 & $1.1\times10^{49}$  & $3.7\times10^{50}$ & $4.5\times10^{50}$ & 0.47 & 5 & 0.11 \\
Bm01 & 5 & Eqt & 1.6 & 10.5 & 0.301 & 0.186 & $1.1\times10^{55}$ & 110 & $1.1\times10^{49}$  & $3.7\times10^{50}$ & $4.5\times10^{50}$ & 0.36 & 14 & 0.082 \\
Bm01 & 10 & Pol & 0.87 & 20 & 0.579 & 0.173 & $4.5\times10^{55}$ & 38 & $3.5\times10^{49}$  & $1.1\times10^{51}$ & $1.4\times10^{51}$ & 0.47 & 7 & 0.11 \\
Bm01 & 10 & Eqt & 1.0  & 19 & 0.579 & 0.173 & $4.5\times10^{55}$ & 38 & $3.5\times10^{49}$  & $1.1\times10^{51}$ & $1.4\times10^{51}$ & 0.40 & 13 & 0.086 \\
Cv03 & 5 & Pol & 1.2 & 12.5 & 0.338 & 0.178 & $1.6\times10^{55}$ & 51 & $1.6\times10^{49}$   & $3.5\times10^{50}$ & $4.9\times10^{50}$ & 0.46 & 8 & 0.12 \\
Cv03 & 5 & Eqt & 1.2 & 10.5 & 0.338 & 0.178 & $1.6\times10^{55}$ & 51 & $1.6\times10^{49}$  & $3.5\times10^{50}$ & $4.9\times10^{50}$ & 0.43 & 10 & 0.10 \\
Dv01 & 5 & Pol & 0.65 & 0.54 & 0.142 & 0.086 & $2.6\times10^{54}$ & 213 & $5.6\times10^{47}$ & $1.8\times10^{49}$ & $2.2\times10^{49}$ & 0.46 & 7 & 0.12 \\
Dv01 & 5 & Eqt & 0.82 & 0.49 & 0.142 & 0.086 & $2.6\times10^{54}$ & 213 & $5.6\times10^{47}$ & $1.8\times10^{49}$ & $2.2\times10^{49}$ & 0.35 & 19 & 0.10 \\
Ev01c1 & 5 & Pol & 0.35 & 0.16 & 0.152 & 0.093 & $9.6\times10^{53}$ & 76 & $2.4\times10^{47}$ & $8.4\times10^{48}$ & $9.6\times10^{48}$ & 0.45 & 6 & 0.13 \\
Ev01c1 & 5 & Eqt & 0.45 & 0.15 & 0.152 & 0.093 & $9.6\times10^{53}$ & 76 & $2.4\times10^{47}$ & $8.4\times10^{48}$ & $9.6\times10^{48}$ & 0.34 & 17 & 0.088 \\
Fo10 & 2 & Pol & 1.4 & 0.26 & 0.118 & 0.188 & $1.2\times10^{53}$ & 238 & $2.2\times10^{47}$ & $3.7\times10^{48}$ & $6.5\times10^{48}$ & 0.52 & 4 & 0.12 \\
Fo10 & 2 & Eqt & 1.7 & 0.28 & 0.118 & 0.188 & $1.2\times10^{53}$ & 238 & $2.2\times10^{47}$ & $3.7\times10^{48}$ & $6.5\times10^{48}$ & 0.44 & 7 & 0.11 \\
Fo10 & 5 & Pol & 0.9 & 0.8 & 0.283 & 0.172 & $7.0\times10^{53}$  & 101  & $1.2\times10^{48}$ & $1.9\times10^{49}$ & $3.6\times10^{49}$ & 0.57 & 2 & 0.13 \\
Fo10 & 5 & Eqt & 1.1 & 0.83 & 0.283 & 0.172 & $7.0\times10^{53}$ & 101 & $1.2\times10^{48}$ & $1.9\times10^{49}$ & $3.6\times10^{49}$ & 0.47 & 5 & 0.11 \\
Fo10 & 11 & Pol & 0.56 & 1.5 & 0.572 & 0.127 & $3.8\times10^{54}$ & 20 & $4.2\times10^{48}$ & $5.2\times10^{49}$ & $1.1\times10^{50}$ & 0.61 & 4 & 0.15 \\
Fo10 & 11 & Eqt & 0.58 & 1.4 & 0.572 & 0.127 & $3.8\times10^{54}$ & 20 & $4.2\times10^{48}$ & $5.2\times10^{49}$ & $1.1\times10^{50}$ & 0.57 & 5 & 0.13 \\
Go60 & 2 & Pol & 1.55 & 4.4 & 0.097 & 0.137 & $5.5\times10^{54}$ & 350 & $2.5\times10^{48}$  & $9.9\times10^{49}$ & $1.1\times10^{50}$ &  0.46 & 5 & 0.12 \\
Go60 & 2 & Eqt & 1.8 & 4.4 & 0.097 & 0.137 & $5.5\times10^{54}$ & 350  & $2.5\times10^{48}$ & $9.9\times10^{49}$ & $1.1\times10^{50}$ & 0.39 & 10 & 0.10 \\
Go60 & 4 & Pol & 0.98 & 6.7 & 0.177 & 0.124 & $1.9\times10^{55}$ & 136 & $6.5\times10^{48}$  & $2.2\times10^{50}$ & $2.5\times10^{50}$ & 0.49 & 5 & 0.13 \\
Go60 & 4 & Eqt & 1.0 & 5.9 & 0.177 & 0.124 & $1.9\times10^{55}$  & 136 & $6.5\times10^{48}$ & $2.2\times10^{50}$ & $2.5\times10^{50}$ & 0.45 & 7 &  0.11 \\
Go60 & 11 & Pol & 0.34 & 3.2 & 0.357 & 0.063 & $7.5\times10^{55}$ & 38 & $1.0\times10^{49}$  & $2.0\times10^{50}$ & $2.5\times10^{50}$ & 0.49 & 5 & 0.16 \\
Go60 & 11 & Eqt & 0.34 & 2.8 & 0.357 & 0.063 & $7.5\times10^{55}$ & 38 & $1.0\times10^{49}$ & $2.0\times10^{50}$ & $2.5\times10^{50}$ & 0.46 & 6 & 0.14 \\
Ho90 & 2  & -- & 1.58 & 6 & 0.088 & 0.122 & $1.0\times10^{55}$ & 402 & $3.4\times10^{48}$  & $1.6\times10^{50}$ & $1.6\times10^{50}$ & 0.50 & 4 & 0.13 \\
Ho90 & 5  & -- & 0.68 & 5.8 & 0.188 & 0.088 & $4.9\times10^{55}$ & 110 & $7.9\times10^{48}$ & $2.9\times10^{50}$ & $2.9\times10^{50}$ & 0.53 & 3 & 0.15 \\
Ho90 & 10 & -- & 0.32 & 3.1 & 0.29 & 0.055 & $1.2\times10^{56}$ & 53 & $1.2\times10^{49}$ & $2.9\times10^{50}$ & $2.9\times10^{50}$ & 0.54 & 3 & 0.15 \\
In15 & 4 & Pol & 1.0 & 3.4 & 0.204 & 0.157 & $4.9\times10^{54}$ & 105 & $3.5\times10^{48}$ & $1.1\times10^{50}$ & $1.4\times10^{50}$ & 0.50 & 5 & 0.11 \\
In15 & 4 & Eqt & 1.2 & 3.1 & 0.204 & 0.157 & $4.9\times10^{54}$ & 105 & $3.5\times10^{48}$ & $1.1\times10^{50}$ & $1.4\times10^{50}$ & 0.40 & 13 & 0.086 \\
Jn20 & 4 & Pol & 1.0 & 3.2 & 0.202 & 0.163 & $4.5\times10^{54}$ & 101 & $3.5\times10^{48}$ & $9.7\times10^{49}$ & $1.4\times10^{50}$ & 0.49 & 6 & 0.10 \\
Jn20 & 4 & Eqt & 1.2 & 2.7 & 0.202 & 0.163 & $4.5\times10^{54}$ & 101 & $3.5\times10^{48}$ & $9.7\times10^{49}$ & $1.4\times10^{50}$ & 0.38 & 15 & 0.071 \\
Knbr & 4 & Pol & 1.3 & 4.7 & 0.198 & 0.144 & $6.5\times10^{54}$ & 163 & $4.2\times10^{48}$ & $1.3\times10^{50}$ & $1.6\times10^{50}$ & 0.46 & 7 & 0.11 \\
Knbr & 4 & Eqt & 1.6 & 4.6 & 0.198 & 0.144 & $6.5\times10^{54}$ & 163 & $4.2\times10^{48}$ & $1.3\times10^{50}$ & $1.6\times10^{50}$ & 0.37 & 16 & 0.088 \\
\hline \hline
\end{tabular}
\label{tab:tab2}
\end{table*}

\subsection{Comparison with Observations}

The growing sample of radio-detected TDEs has revealed two broadly distinct classes of radio emission behavior \citep{cendes2024}. The first encompasses flares that rise to a peak within a few hundred days after the optical discovery \citep{alexander2020}. The early-time flares may correspond to the steeper CNM density profiles ($n\geq 2$, see runs Jn20 in Fig. \ref{fig7} and Nn25 in Fig. \ref{figA4}), where the rapidly declining ambient density causes the light curve to peak on short timescales. The second class comprises flares that emerge more than $\sim$1000 days after discovery. The physical origin of the latter remains under active debate. 

Regarding a direct comparison with observed light curves, we suggest that the actual situation is more complex than a straightforward comparison would imply. 

\textit{First, the definition of the time zero point.} Our theoretical light curves adopt the outflow launch epoch as $t = 0$, whereas observational studies universally adopt the optical TDE discovery time as $t = 0$. If the outflow is indeed delayed relative to the optical burst, and the time zero point is moved to the delayed outflow launch time, the observed rise rate would become considerably shallower, and more consistent with the gradual rise ($\Gamma_1 \lesssim 1$) as expected from the FS scenario. Accordingly, a subset of the light curves reported in the observational literature cannot be directly compared to our results without first correcting for this zero-point offset. 

\textit{Second, genuine mismatches as a scientific result.} A subset of observed radio TDEs exhibit light curves that are difficult to reconcile with our simulation results. These include sources displaying two distinct luminosity peaks separated by several years (e.g., ASASSN-15oi, AT2019dsg in \citealt{cendes2024}), as well as sources undergoing rapid flux variations on sub-year timescales. As discussed above, sources that are inconsistent with pure FS predictions imply that additional physical mechanisms are at play. Candidate explanations include the bow shock scenario (which we investigate in a companion paper \citealt{mou2025b}) and delayed secondary outflows \citep{horesh2021}. 

The primary goal of this paper is to establish the fundamental radio emission characteristics expected from the FS scenario across a wide parameter space, providing a necessary reference before applying specific models to individual sources. A detailed and quantitative comparison between simulations and observed radio light curves is an important next step, which we plan to address in future work.

\section{Conclusions}  

In this study, we conducted hydrodynamic simulations to explore the FS scenario of the radio afterglows frequently observed in TDEs. 
We inject the CRe component into the grids near the shock front and simulate its hydrodynamic evolution after the shock acceleration. 
Based on the hydrodynamic data, we calculated the emitting radio spectra along two different directions using radiative transfer. 

We demonstrate that within a CNM environment similar to that of the Galactic Center, a moderate outflow (with kinetic energies around $10^{50}$ erg) can produce radio afterglows at the mJy level for $d_L=100$ Mpc, with the self-absorption frequency near the GHz range. The corresponding luminosity is on the order of $10^{37}~\ergs$. Although the flux depends on the outflow strength and CNM density, we find that fluxes above the 10 mJy level require either a very strong outflow ($10^{51}$ erg) or a density significantly higher than that of the Galactic Center. 

The self-absorption frequency $\nu_p$ shows a two-stage power-law evolution, in which $\nu_p$ decreases slowly in the early stage and rapidly in the late stage. 

We find that the synthetic radio spectra exhibit an anisotropic feature that has not been previously reported. In particular, when the outflow opening angle is small, the radio spectrum in the equatorial direction shows a ``flat-top'' feature. 

Previous observational studies commonly employ analytical formulas of the equipartition method to estimate the shock parameters. 
After taking into account the specific distribution of CRe and radiative transfer, we reveal a significant discrepancy between the analytically estimated values and the actual values: (1) the energies of CRe in simulations ($E_e$ in Table \ref{tab:tab2}) can be, on average, as much as 9-10 times higher than the estimated ones ($E'_e$ in Table \ref{tab:A1}); (2) the shock energies ($E_s$) are on average 9 times higher than the estimated ones ($E'_s$); (3) the shock radii are twice of those estimated values; (4) the post-shock CNM densities ($n_e$) are lower by a factor of several compared to the estimated values ($n'_e$).

Under most parameter settings, the flux exhibits an initial rising phase followed by a subsequent decline phase. 
An exception occurs when the density index $n > 1.5$: after a brief rise lasting several months, only the declining phase remains. 
The radio flux rises relatively slowly (flatter than $t^{1.0}$), and its decay is also gradual, with the flux halving timescale of several years. 
These variation features can be used to test the applicability of the FS scenario, which are quite different from the bow shock scenario, in which the radio flux can exhibit a sharp rise followed by a rapid decline (\citealt{mou2022, zhuang2025}). 
We also find that when the high-frequency radio flux begins to decline, the shock energy at that turnover moment is approximately one half of the outflow's kinetic energy.

Finally, we would like to emphasize that there are several uncertainties in the interaction between the outflow and the CNM, such as the angular distribution of the outflow, the possibility of abrupt changes in the radial distribution of the hot/diffuse CNM, and the evolution of the magnetic field. 
Given the current limited constraints on the physics of the outflow, CNM and shock, our results are based on simplified assumptions. 
Further progress in understanding the radio afterglows will necessitate a coordinated development between theoretical modeling and observations.


\section*{Acknowledgements}
We sincerely thank the referee for the insightful comments. 
G.M. was supported by the National Key R\&D Program of China (Grant No. 2023YFA1607904), and the NSFC (No. 12473013).

\appendix
\setcounter{figure}{0} 
\renewcommand{\thefigure}{A\arabic{figure}}
\setcounter{table}{0}
\renewcommand{\thetable}{A\arabic{table}}

\section{The influence of the light-travel time difference}
\begin{figure}
\centering
 \includegraphics[width=0.48\columnwidth]{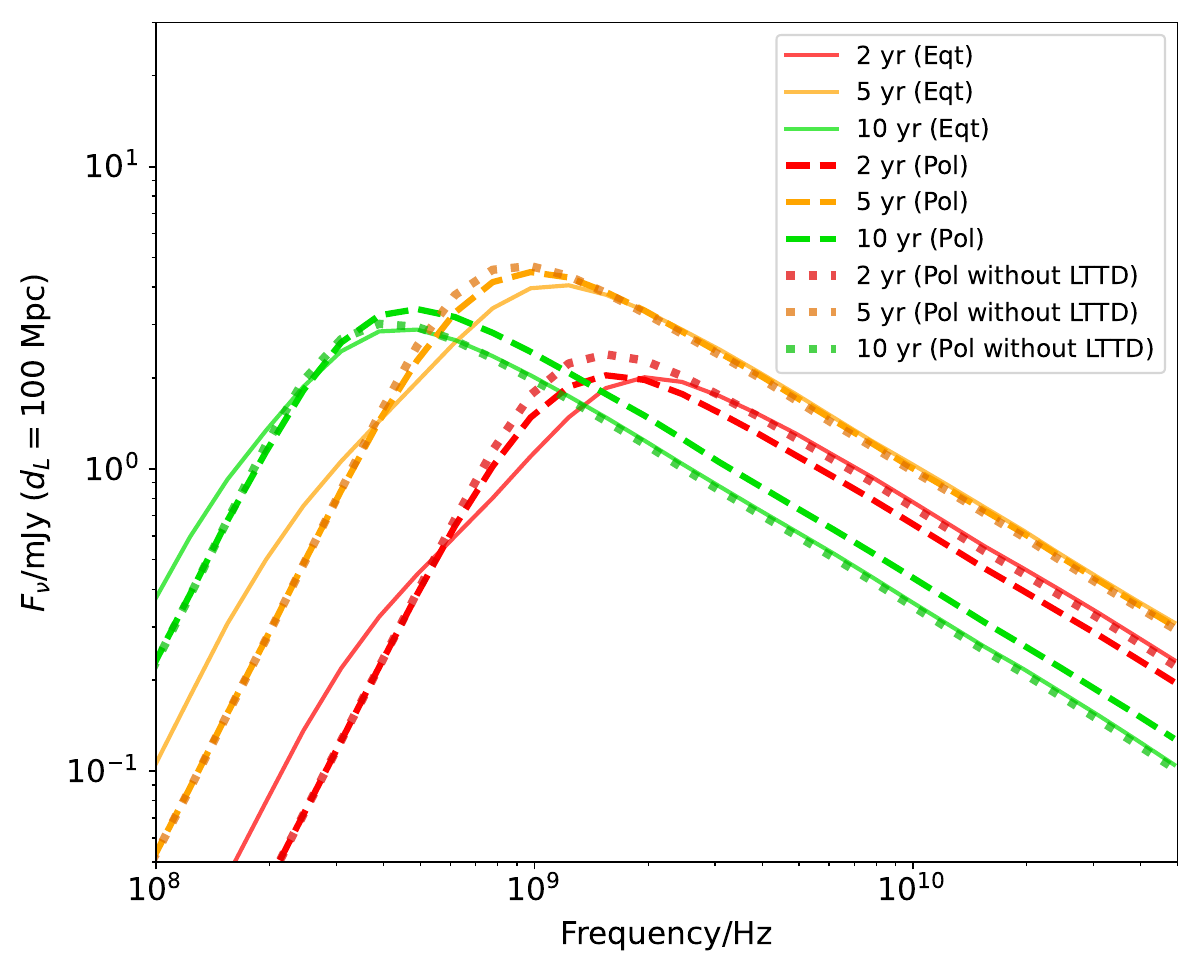}
 \caption{Comparison of the spectra in the polar direction with and without considering light-travel time difference (LTTD) under the fiducial model (run A). The results suggest that the inclusion of LTTD could slightly affect the spectra, as evidenced by the comparison between the dashed and dotted lines with the same colors. } 
 \label{figA1}
\end{figure}

We show in Figure \ref{figA1} the impact of considering versus neglecting LTTD on the spectra along the polar direction for run A. The results indicate that LTTD can affect the emergent spectrum in the polar direction, although the effect is not particularly significant. In the main text, all spectral calculations along the polar direction have taken LTTD into account.

The non-coincidence of the radio spectra at high frequencies from two different viewing angles is caused jointly by the LTTD and the temporal evolution of the high-frequency flux. Consider two shocks departing from the origin and moving outward along the two polar axes with a velocity $v$. An observer located in the polar direction simultaneously receives the radiation emitted by the near-side shock at an age of $t_1$ and the radiation emitted by the far-side shock at an age of $t_2$. Then, $t_1$ and $t_2$ satisfy the relation:
\begin{equation}
t_1 - t_2 = \frac{v t_1 + v t_2}{c},
\end{equation}
which yields:
$t_2 = t_1 \frac{c - v}{c + v} $.

Assume that in the source region, the evolution of the high-frequency flux $F_{\nu}$ with time follows the relation $F_{\nu} \propto t^{\Gamma}$, where $\Gamma$ varies over time. The total flux in the polar direction is:
\begin{equation}
F_{\mathrm{pol}} = t_1^{\Gamma} + t_2^{\Gamma} = t_1^{\Gamma} \left[ 1 + \left(1 - \frac{2v}{v+c}\right)^{\Gamma} \right].
\end{equation}

In contrast, the equatorial direction does not take LTTD into account, so the high-frequency flux is given by:
$F_{\mathrm{eqt}} = 2 t_1^{\Gamma}$. 
Consequently, the ratio of the high-frequency flux between the two directions is:
\begin{equation}
\frac{F_{\rm pol}} {F_{\rm eqt}}= \dfrac{1 + (1 - \frac{2v}{c+v})^{\Gamma}}{2}
\end{equation}

In our simulation, $v = 0.2c$, which simplifies the above ratio to $F_{\rm pol}/F_{\rm eqt}=(1 + 0.67^{\Gamma})/2$.
This is a rough estimate, since $v$ would decrease in the late stage. 
Since $\Gamma$ decreases over time (Section 5.3), we can observe the following behavior for the high-frequency flux in both directions: they differ significantly in the early phase with $F_{\mathrm{pol}} < F_{\mathrm{eqt}}$ (where $\Gamma \simeq 1.1$), are close to each other in the middle stage ($\Gamma \rightarrow 0$), and reverse in the late stage such that $F_{\mathrm{pol}} > F_{\mathrm{eqt}}$ (where $\Gamma < 0$).

\section{The usage procedure of the energy equipartition/minimal method} \label{app2}
\citet{duran2013} developed the energy equipartition/minimal method for synchrotron emission, which provides estimate for the size of the radiation zone, and the energies of magnetic field and CRe. However, some observational articles failed to apply this method correctly, resulting in that first, the number of CRe components was mistakenly regarded as the whole gas including thermal components (as pointed out by \citealt{matsumoto2021}), and second, the derived magnetic field energy and CRe energy were inconsistent with the specified values of $\epb$ and $\epe$.  

The reason for the second point is that, many observational studies implicitly employ the three--component minimal energy method when a factor $\xi^{11/(13+2p)}$ is applied to the $E_{\rm eq}$ calculation formula ($\xi \equiv 1+ \epe^{-1}$) (equation 28 in \citealt{duran2013}, see Section 4.2.2 therein for details). In this case, $E_{\rm eq}$ actually is $E_e + E_B +  E_{\rm thp}$ ($E_{\rm thp}$ is the total energy of thermal protons), in which these three components satisfy $E_e: E_B: E_{\rm thp}= \frac{11}{17}\xi^{-1} : \frac{6}{17}: \frac{11}{17}(1-\xi^{-1})$. In other words, once $\xi$ or $\epe$ is specified, the relationship among $E_B$, $E_e$, and $E_{\rm thp}$ is already determined, which is therefore inconsistent with further specifying the value of $\epb$. As a result, the values of $E_e$ and $E_B$ reported in many studies often do not match their adopted values of $\epe$ and $\epb$. 

The procedure of the equipartition method should follow the way to find equipartition radius $R_{\rm eq}$ when ($E_e+E_B$) reaches the minimal value, and then derive the values under the specified value of assumed $\epb$ and $\epe$. Following this procedure, each calculation step has well-defined physical meaning. All the following parameters in Step 1 -- 3 are the same as \citet{duran2013}, and all the parameters in the formula are unchanged.

{\bf Step 1: Calculating the equipartition radius $R_{\rm eq}$. } 

When $E_B= (6/11) E_e$, $E_e+E_B$ reaches the minimal value. In this case, the radius of the radiation zone is also determined. 
\begin{eqnarray} \label{Req} 
R_{\rm eq} &\approx& (1\times10^{17} {\rm cm}) \, [21.8(525)^{p-1}]^{\frac{1}{13+2p}} \,
\gamma_m^{\frac{2-p}{13+2p}} \, \nonumber \\
&&
\times ~ \Big[ F_{\rm p,mJy}^{\frac{6+p}{13+2p}} \, d_{L,28}^{\frac{2(p+6)}{13+2p}}
\, \nu_{p,10}^{-1} \, (1+z)^{-\frac{19+3p}{13+2p}} \Big] \,  \nonumber \\
&&
\times ~ f_A^{-\frac{5+p}{13+2p}} \, f_V^{-\frac{1}{13+2p}}  \,
\Gamma^{\frac{p+8}{13+2p}} \,  4^{\frac{1}{13+2p}} . 
\end{eqnarray}
This is just equation 27 in \citep{duran2013} except for an extra correction term $4^{\frac{1}{13+2p}}$ for the isotropic number of radiating electrons. In this formula, $\gamma_m = \chi_e (\Gamma-1)$, where $\chi_e = \frac{p-2}{p-1}\epsilon_e \frac{m_p}{m_e}$, and if 
$\gamma_m$ is found to be $\gamma_m < 2$, one should use $\gamma_m = 2$. 
Note the value of $\Gamma$ is the bulk Lorentz factor of the radiation zone, and one can set it to be $\Gamma=1$ in Step 1--3 before iteration. 

{\bf Step 2: Calculating $R$, $E_e$, $E_B$, $N_e$ and $B$. }

After setting the values of $\epb$ and $\epe$, the condition for the minimal energy is usually broken, and the radius of the radiation zone slightly deviates from $R_{\rm eq}$ by a factor of $\epsilon^{1/17}$: 
\be
~~~ R = R_{\rm eq} \epsilon^{1/17}  ~,
\ee
where $\epsilon \equiv 11\epb/(6\epe)$.  
The energy in relativistic electrons within the observed region with $\gme \geq \gamma_m$ is
\begin{eqnarray} \label{E_e}
E_e &=& N_e m_e c^2 \gamma_e \Gamma = \frac{4 (\gamma_e/\gamma_m)^{2-p} \, 27 c^3 F_{\nu,p}^4 d_L^8 \eta^5 \Gamma^2}{16 \sqrt{3} \pi^3 e^2
  m_e^2 \nu_p^7 (1+z)^{11} f_A^3 R^6} \nonumber \\
&\approx&  4 (\gamma_e/\gamma_m)^{2-p} \times 4.4 \times 10^{50} {\rm erg}  \times ~ \Big[ F_{\rm p,mJy}^4 \, d_{L,28}^8 \, \nu_{p,10}^{-7} \, \eta^{5}
\, (1+z)^{-11} \Big] \, \frac{ \Gamma^2}{f_A^{3} \, R_{17}^{6}},
\end{eqnarray}
while the energy in the magnetic field is
\begin{eqnarray}
E_B &=& \frac{(B \Gamma)^2}{8 \pi} V =  \frac{8 \pi^6 m_e^6 c^2 \nu_p^{10}
  (1+z)^{14} f_A^4 f_V R^{11}}{81 e^2 F_{\nu,p}^4 d_L^8 \eta^{\frac{20}{3}} \Gamma^8} \nonumber \\
&\approx&  (2.1 \times 10^{46} {\rm erg}) \, \left[  F_{\rm p,mJy}^{-4} \, d_{L,28}^{-8} \, \nu_{p,10}^{10} \, \eta^{-\frac{20}{3}}
\, (1+z)^{14}  \right] \, 
\times ~ \frac{f_A^{4} \, f_V \, R_{17}^{11}}{ \Gamma^{8}},
\end{eqnarray}
where the volume of the radiation zone $V = f_V \pi R^3 /\Gamma^4$. 
These are equation 17 and 18 in \citep{duran2013}, except for an extra correction term $4(\gme/\gamma_m)^{2-p}$ in equation \ref{E_e} for the isotropic number of radiating electrons and those electrons with $\gme \geq \gamma_m$. 

Subsequently, one can derive the values of $B$ and $N_e$ from the values of $E_e$, $E_B$ and $R$. 
\begin{eqnarray} \label{N_e}
N_e &=& \frac{4 (\gamma_e/\gamma_m)^{2-p} \, 9 c F_{\nu,p}^3 d_L^6 \eta^{\frac{10}{3}}}{8 \sqrt{3} \pi^2 e^2 m_e^2 \nu_p^5 (1+z)^8 f_A^2 R^4} \nonumber \\
&\approx& 4 \times 10^{54} \, (\gamma_e/\gamma_m)^{2-p} \,  \left[ F_{\rm p,mJy}^3 \, d_{L,28}^6 \, \nu_{p,10}^{-5} \,
\eta^{\frac{10}{3}} \,(1+z)^{-8} \right] \, 
\times ~ \frac{1}{ f_A^{2} \, R_{17}^{4}},
\end{eqnarray}
\begin{eqnarray}
B &=&  \frac{8 \pi^3 m_e^3 c \nu_p^5 (1+z)^7 f_A^2 R^4}{9 e F_{\nu,p}^2 d_L^4 \eta^{\frac{10}{3}} \Gamma^3} \nonumber \\
&\approx& (1.3 \times 10^{-2} \, {\rm G}) \, \left[ F_{\rm p,mJy}^{-2} \, d_{L,28}^{-4} \, \nu_{p,10}^5 \,
\eta^{-\frac{10}{3}} \, (1+z)^7 \right] \, 
\times ~ \frac{f_A^{2} \, R_{17}^4}{\Gamma^{3}}.
\end{eqnarray}
These are equation 15 and 16 in \citep{duran2013}, except for the an extra correction term $4(\gme/\gamma_m)^{2-p}$ in equation \ref{N_e}. Note that $N_e$ is the number of relativistic electrons with $\gme \geq \gamma_m$, which does not include the thermal components. Thus, one should not use $N_e$ directly to calculate the density of hot CNM. 
The Lorentz factor of the relativistic electrons radiating at $\nu_p$ is (equation 14 in \citealt{duran2013}): 
\begin{eqnarray} \label{gamma_a}
\gamma_a &=& \frac{3 F_{\nu,p} d_L^2 \eta^{\frac{5}{3}} \Gamma}{2 \pi \nu_p^2 (1+z)^3 m_e f_A R^2 } 
\approx 525 \, \left[F_{\rm p,mJy} \, d_{L,28}^2 \, \nu_{p,10}^{-2} \, 
\eta^{\frac{5}{3}} \, (1+z)^{-3}\right] \, \frac{\Gamma}{f_A \, R_{17}^{2}} .
\end{eqnarray} 

{\bf Step 3: Finding out $R_s(t)$ and checking the self-consistency.  }

According to the radio data achieved in different periods, one can derive the relationship between shock radius and time: $R_s(t)$, from which the shock velocity $v_s$ and $\Gamma -$value can be obtained. 
Subsequently, do iterations and bring the $\Gamma -$value back to equations \ref{Req}--\ref{gamma_a} in Step 1--2 to obtain the parameter values.

{\bf Step 4: Calculating the shock energy and the CNM density.} 

The shock energy is calculated as $E_{s}= E_e / \epsilon_e$.  
The enclosed mass of the hot CNM swept up by the FS can be derived as
$M_{\rm cnm}(R_s)=E_s/[(\Gamma-1)c^2]$ in the relativistic regime, or $
M_{\rm cnm}(R_s)=2E_s/v^2_s $ in the nonrelativistic regime. 

The averaged post-shock CNM density can be estimated by $\rho (R_s) \simeq M_{\rm cnm}(R_s)/V$, while the density of the the pre-shock CNM ahead of the shock front should be further reduced by a factor of 4 (density jump). 
It should be noted that the shock energy $E_s$ is not equivalent to the outflow energy $E_k$; rather, $E_s$ represents a robust lower limit for $E_k$.  
When the swept-up CNM mass is significantly less than the outflow mass, the outflow energy can be approximated by
\be
E_k \simeq E_s \times \frac{M_{\rm out}}{M_{\rm cnm}}
\ee

Due to the lack of constraints on $M_{\rm out}$, it is challenging to estimate $E_k$, and it may be approximated as $E_s$ when the forward shock velocity decreases significantly for the FS scenario. 

A program implementing the above procedures is available on GitHub \footnote{https://github.com/G-Mou/RadioTDE} and Zenodo (DOI: 10.5281/zenodo.20268807). 
Please note: The above program and web-based code are intended only to illustrate the workflow of the energy equipartition method, and do not account for the errors introduced by this approach. One should note that the equipartition method provides \emph{rough} estimates of the shock radius, $E_e$ and $E_B$, since the derivation in \citet{duran2013} involves multiple approximations, and could lead to considerable deviation. 
For better matching the results of radiation transfer calculations, one should multiply the $R_{\rm eq}$ by a factor of 2, $E_e$ and $E_B$ by a factor of 9 (on average).

\section{Shock parameters derived from the equipartition method }
Based on the synthetic radio spectra obtained from the hydrodynamic simulations, we extract $\nu_p$ and $F_{\nu_p}$, and then follow the steps of the equipartition method to derive the ``estimated'' shock radius $R_{\rm eq}$, density $n'_e$, and various energy components ($E'_e, E'_B, E'_s$) in Table \ref{tab:A1}. These parameters are used for comparison with the actual values in simulations (Table \ref{tab:tab2}). 
 
\begin{table*}[htbp]
\centering
\scriptsize
\caption{Shock parameters derived by using the equipartition method based on the synthetic radio spectra. We assume that the emitting region is a shell of thickness 0.1$R_{\rm eq}$, and opening angle of the shock is equal to the outflow's opening angle in each model. }
\begin{tabular}{ccccccccccccc}
\hline
\hline
 Run &  $\Delta T$ & Direc & $\nu_p$ & $F_{\nu_p}$ & $R_{\rm eq}$ & $V^{'}_{\rm s}\equiv R_{\rm eq}/\Delta T$ & $N'_e$ & $n'_e$ & $E'_e$ & $E'_B$ & $E'_{\rm s}$ \\
 (units) & yr & -- & GHz  & mJy & pc & c &   & $\cmc$ & erg & erg & erg  \\
\hline
A & 2   & Pol           & 1.6 & 2  & 0.049      & 0.080 & $9.1\times10^{53}$ & 1762 & $1.5\times10^{47}$ & $5.8\times10^{47}$ & $5.2\times10^{48}$ \\
A  & 2  & Eqt           & 2.0 & 2.0  & 0.039   & 0.064 & $1.1\times10^{54}$ & 4300 & $1.2\times10^{47}$ & $4.6\times10^{47}$ & $4.1\times10^{48}$ \\
A  & 2  & Eqt ($p2.2$) & 2.3 & 3.1 & 0.042 & 0.069 & $7.7\times10^{53}$ & 2870 & $9.8\times10^{46}$ & $3.4\times10^{47}$ & $3.3\times10^{48}$ \\
A  & 2  & Eqt ($p2.8$)& 1.7 & 1.1 & 0.035  & 0.057  & $1.5\times10^{54}$ & 6644 & $1.3\times10^{47}$ & $5.0\times10^{47}$ & $4.2\times10^{48}$ \\
A  & 5  & Pol            & 0.98 & 4.4 & 0.116 & 0.076 & $4.2\times10^{54}$ & 615 & $6.5\times10^{47}$ & $2.4\times10^{48}$ & $2.2\times10^{49}$ \\
A  & 5  & Eqt           & 1.1 & 4.1 & 0.099 & 0.065    & $4.7\times10^{54}$ & 1094 & $5.3\times10^{47}$ & $2.0\times10^{48}$ & $1.8\times10^{49}$ \\
A  & 5  & Eqt ($p2.2$) & 1.3 & 6.2 & 0.103 & 0.067  & $3.3\times10^{54}$ & 829 & $3.9\times10^{47}$ & $1.4\times10^{48}$ & $1.3\times10^{49}$ \\
A  & 5  & Eqt ($p2.8$) & 1.0 & 2.3 & 0.084 & 0.055 & $6.5\times10^{54}$ & 2107 & $5.2\times10^{47}$ & $2.1\times10^{48}$ & $1.7\times10^{49}$ \\
A  & 10 & Pol           & 0.48 & 3.4 & 0.21 & 0.069   & $7.7\times10^{54}$ & 189 & $9.7\times10^{47}$ & $3.6\times10^{48}$ & $3.2\times10^{49}$ \\
A  & 10 & Eqt          & 0.46 & 2.9 & 0.203 & 0.066 & $7.3\times10^{54}$ & 197 & $8.4\times10^{47}$ & $3.1\times10^{48}$ & $2.8\times10^{49}$ \\
A  & 10 & Eqt ($p2.2$) & 0.52 & 4.5 & 0.221 & 0.072 & $4.9\times10^{54}$ & 124 & $6.7\times10^{47}$ & $2.3\times10^{48}$ & $2.2\times10^{49}$ \\
A  & 10 & Eqt ($p2.8$) & 0.4 & 1.7 & 0.182 & 0.060 & $9.8\times10^{54}$ & 313 & $9.1\times10^{47}$ & $3.6\times10^{48}$ & $3.0\times10^{49}$ \\
Bm01 & 5 & Pol          & 1.25 & 10.7 & 0.140 & 0.092 & $6.5\times10^{54}$ & 560 & $1.5\times10^{48}$ & $5.5\times10^{48}$ & $4.9\times10^{49}$ \\
Bm01 & 5 & Eqt          & 1.6 & 10.5 & 0.108 & 0.071 & $8.4\times10^{54}$ & 1547 & $1.1\times10^{48}$ & $4.2\times10^{48}$ & $3.7\times10^{49}$ \\
Bm01 & 10 & Pol         & 0.87 & 20.0 & 0.270 & 0.088 & $2.2\times10^{55}$ & 258 & $4.5\times10^{48}$ & $1.7\times10^{49}$ & $1.5\times10^{60}$ \\
Bm01 & 10 & Eqt         & 1.0 & 19.0 & 0.230 & 0.075 & $2.4\times10^{55}$ & 475 & $3.7\times10^{48}$ & $1.4\times10^{49}$ & $1.2\times10^{50}$ \\
Cv03 & 5 & Pol          & 1.2 & 12.5 & 0.157 & 0.103 & $6.5\times10^{54}$ & 398 & $1.8\times10^{48}$ & $6.8\times10^{48}$ & $6.1\times10^{49}$ \\
Cv03 & 5 & Eqt          & 1.2 & 10.5 & 0.145 & 0.095 & $6.2\times10^{54}$ & 487 & $1.5\times10^{48}$ & $5.5\times10^{48}$ & $5.0\times10^{49}$ \\
Dv01 & 5 & Pol          & 0.65 & 0.54 & 0.065 & 0.042 & $1.7\times 10^{54}$ & 1409 & $8.0\times10^{46}$ & $3.0\times10^{47}$ & $2.7\times10^{48}$ \\
Dv01 & 5 & Eqt          & 0.82 & 0.49 & 0.049 & 0.032 & $2.1\times 10^{54}$ & 3947 & $5.7\times10^{46}$ & $2.1\times10^{47}$ & $1.9\times10^{48}$ \\
Ev01c1 & 5 & Pol         & 0.35 & 0.16 & 0.069 & 0.045 & $6.2\times10^{53}$ & 446 & $3.5\times10^{46}$ & $1.3\times10^{47}$ & $1.2\times10^{48}$ \\
Ev01c1 & 5 & Eqt         & 0.45 & 0.15 & 0.052 & 0.034 & $7.7\times10^{53}$ & 1293 & $2.5\times10^{46}$ & $9.3\times10^{46}$ & $8.4\times10^{47}$ \\
Fo10 & 2 & Pol         & 1.4 & 0.26 & 0.061 & 0.10 & $8.7\times10^{52}$ & 860 & $2.3\times10^{46}$ & $8.6\times10^{46}$ & $7.7\times10^{47}$ \\
Fo10 & 2 & Eqt          & 1.7 & 0.28 & 0.052 & 0.085 & $1.1\times10^{53}$ & 1723 & $2.1\times10^{46}$ & $7.7\times10^{46}$ & $6.9\times10^{47}$ \\
Fo10 & 5 & Pol          & 0.9 & 0.8 & 0.161 & 0.105 & $4.6\times10^{53}$ & 246 & $1.4\times10^{47}$ & $5.1\times10^{47}$ & $4.6\times10^{48}$ \\
Fo10 & 5 & Eqt        & 1.1 & 0.83 & 0.134 & 0.088 & $5.7\times10^{53}$ & 529 & $1.2\times10^{47}$ & $4.4\times10^{47}$ & $3.9\times10^{48}$ \\
Fo10 & 11 & Pol        & 0.56 & 1.5 & 0.347 & 0.103 & $1.6\times10^{54}$ & 86 & $4.7\times10^{47}$ & $1.7\times10^{48}$ & $1.6\times10^{49}$ \\
Fo10 & 11 & Eqt          & 0.58 & 1.4 & 0.324 & 0.096 & $1.7\times10^{54}$ & 108 & $4.2\times10^{47}$ & $1.5\times10^{48}$ & $1.4\times10^{49}$ \\
Go60 & 2 & Pol        & 1.55 & 4.4 & 0.045 & 0.074 & $2.6\times10^{54}$ & 1781 & $3.8\times10^{47}$ & $1.4\times10^{48}$ & $1.3\times10^{49}$ \\
Go60 & 2 & Eqt        & 1.8 & 4.4 & 0.038 & 0.062 & $3.2\times10^{54}$ & 3419 & $3.3\times10^{47}$ & $1.2\times10^{48}$ & $1.1\times10^{49}$ \\
Go60 & 4 & Pol         & 0.98 & 6.7 & 0.086 & 0.070 & $7.6\times10^{54}$ & 724 & $9.9\times10^{47}$ & $3.3\times10^{48}$ & $3.3\times10^{49}$ \\
Go60 & 4 & Eqt         & 1.0 & 5.9 & 0.079 & 0.065 & $7.4\times10^{54}$ & 900 & $8.3\times10^{47}$ & $2.8\times10^{48}$ & $2.8\times10^{49}$ \\
Go60 & 11 & Pol        & 0.34 & 3.2 & 0.175 & 0.052 & $1.6\times10^{55}$ & 186 & $1.2\times10^{48}$ & $3.9\times10^{48}$ & $3.9\times10^{49}$ \\
Go60 & 11 & Eqt        & 0.34 & 2.8 & 0.164 & 0.049 & $1.6\times10^{55}$ & 216 & $1.0\times10^{48}$ & $3.4\times10^{48}$ & $3.4\times10^{49}$ \\
Ho90 & 2    & -- & 1.58 & 6.0 & 0.044 & 0.072 & $4.3\times10^{44}$ & 1624 & $6.0\times10^{47}$ & $2.2\times10^{48}$ & $2.0\times10^{49}$ \\
Ho90 & 5    & -- & 0.68 & 5.8 & 0.100 & 0.065 & $1.2\times10^{55}$ & 370 & $1.3\times10^{48}$ & $4.9\times10^{48}$ & $4.4\times10^{49}$ \\
Ho90 & 10   & -- & 0.32 & 3.1 & 0.158 & 0.052 & $1.9\times10^{55}$ & 148 & $1.3\times10^{48}$ & $5.0\times10^{48}$ & $4.5\times10^{49}$ \\
In15 & 4 & Pol           & 1.0 & 3.4 & 0.102 & 0.083 & $2.5\times10^{54}$ & 568 & $4.7\times10^{47}$ & $1.7\times10^{48}$ & $1.6\times10^{49}$ \\
In15 & 4 & Eqt          & 1.2 & 3.1 & 0.081 & 0.066 & $3.0\times10^{54}$ & 1320 & $3.5\times10^{47}$ & $1.3\times10^{48}$ & $1.2\times10^{49}$ \\
Jn20 & 4 & Pol          & 1.0 & 3.2 & 0.099 & 0.081 & $2.4\times10^{54}$ & 605 & $4.3\times10^{47}$ & $1.6\times10^{48}$ & $1.4\times10^{49}$ \\
Jn20 & 4 & Eqt          & 1.2 & 2.7 & 0.076 & 0.062 & $2.8\times10^{54}$ & 1542 & $3.0\times10^{47}$ & $1.1\times10^{48}$ & $9.9\times10^{48}$ \\
Knbr & 4 & Pol          & 1.3 & 4.7 & 0.091 & 0.074 & $2.5\times10^{54}$ & 1124 & $5.3\times10^{47}$ & $2.0\times10^{48}$ & $1.8\times10^{49}$ \\
Knbr & 4 & Eqt          & 1.6 & 4.6 & 0.073 & 0.060 & $3.0\times10^{54}$ & 2601 & $4.2\times10^{47}$ & $1.6\times10^{48}$ & $1.4\times10^{49}$ \\
\hline
\end{tabular}
\label{tab:A1}
\end{table*}

\begin{figure*}
\centering
\includegraphics[width=0.45\columnwidth]{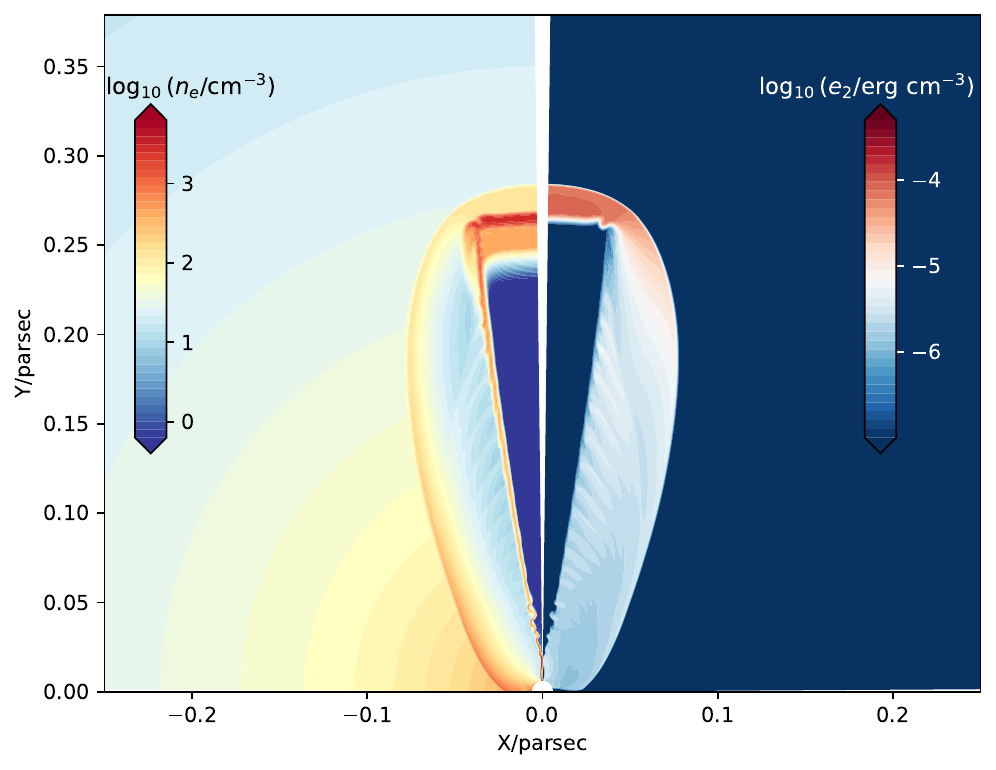}
\includegraphics[width=0.45\columnwidth]{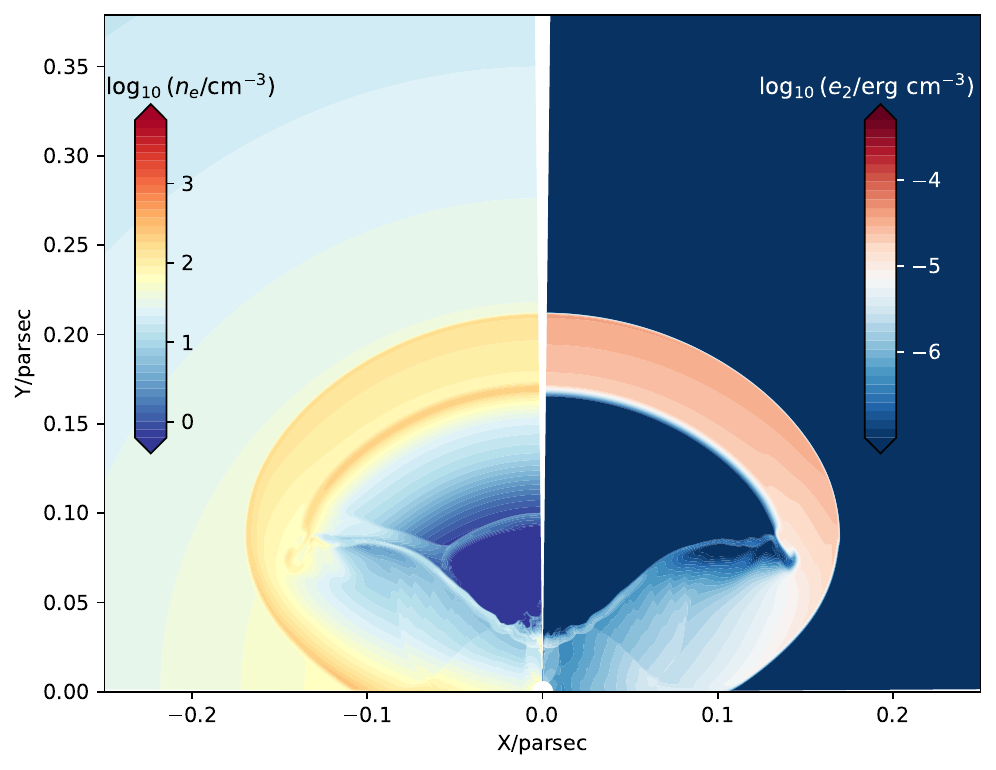}
 \caption{ Snapshots of run Fo10 and Go60 at $t=5$ yr. } 
 \label{figA2}
\end{figure*}

\begin{figure*}
\includegraphics[width=0.5\columnwidth]{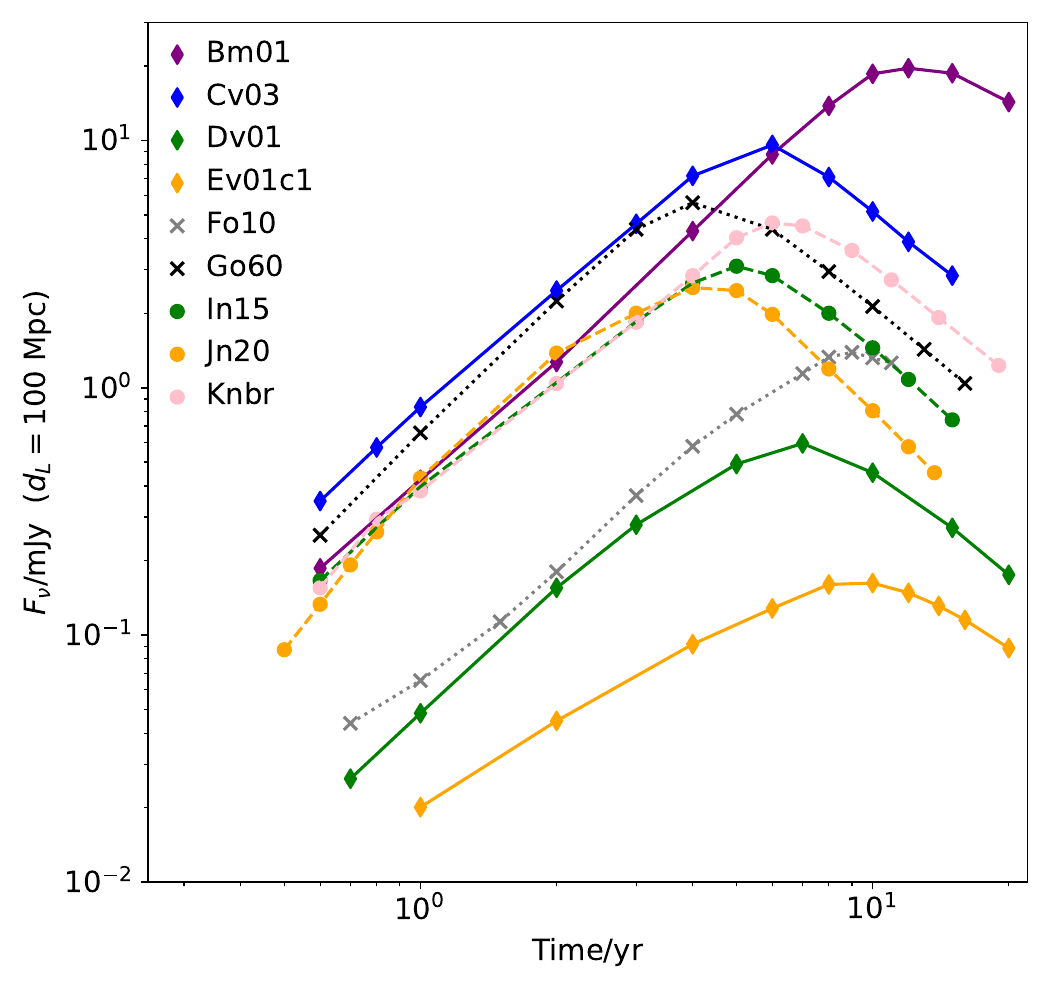}
\includegraphics[width=0.5\columnwidth]{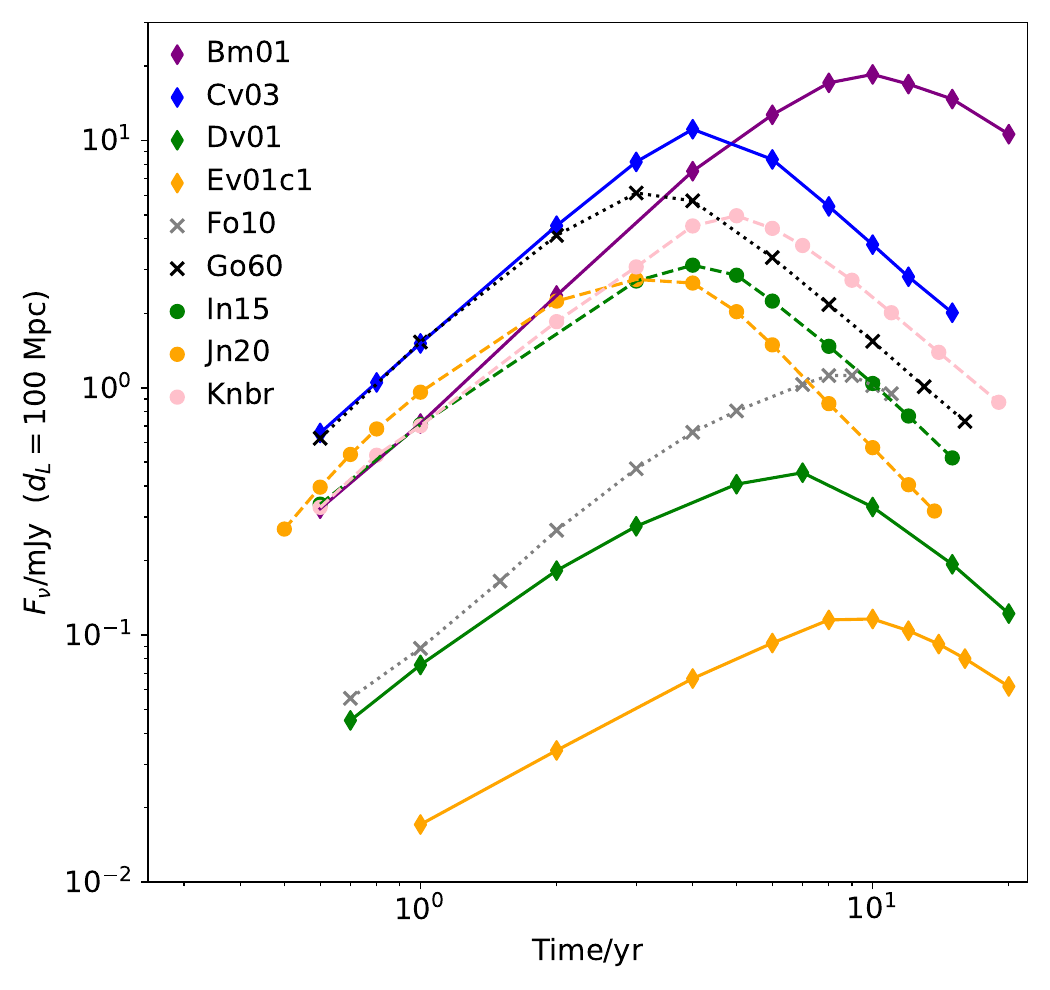}
 \caption{Temporal evolution of the monochromatic flux $F_{\nu}$ in the equatorial direction. The left panel is for $\nu=0.88$ GHz, and the right panel is for 1.4 GHz.  } 
\label{figA3}
\end{figure*}

\begin{figure*}
\includegraphics[width=0.5\columnwidth]{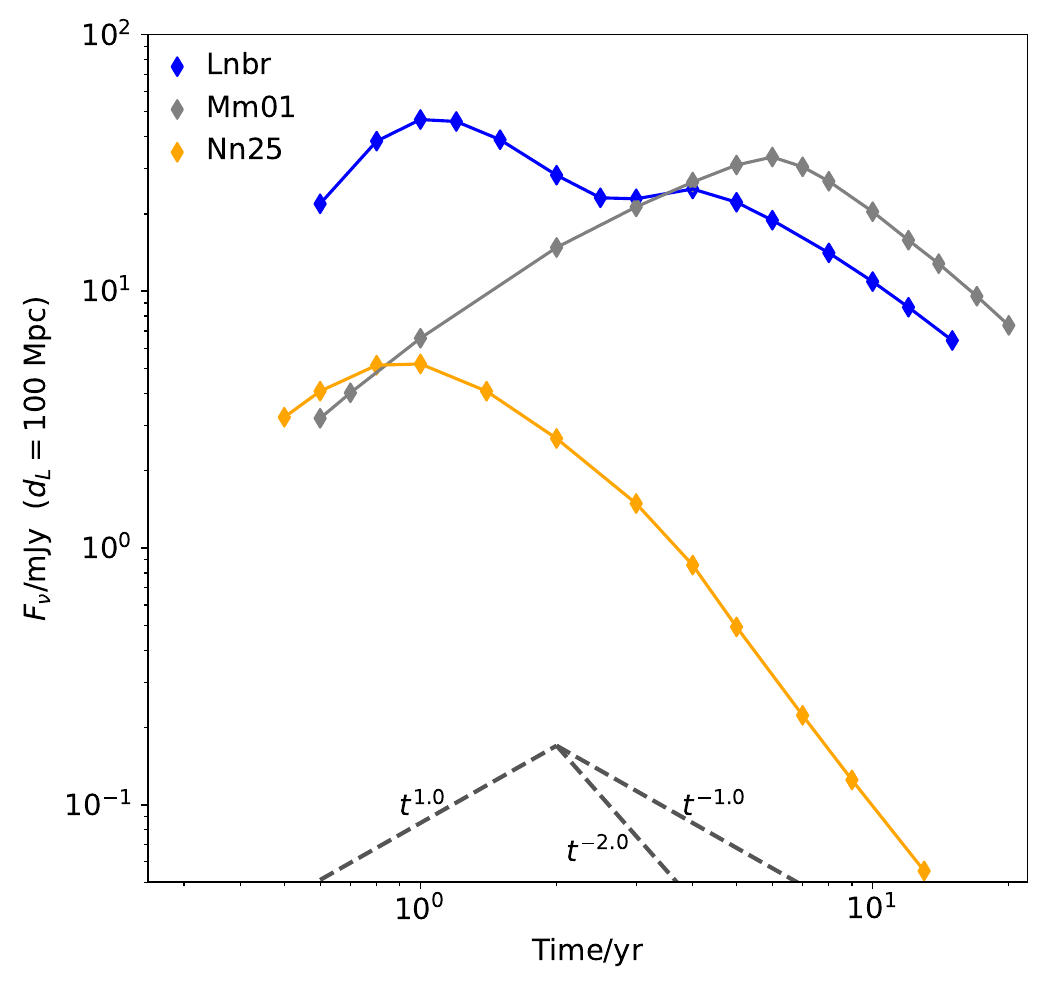}
\includegraphics[width=0.5\columnwidth]{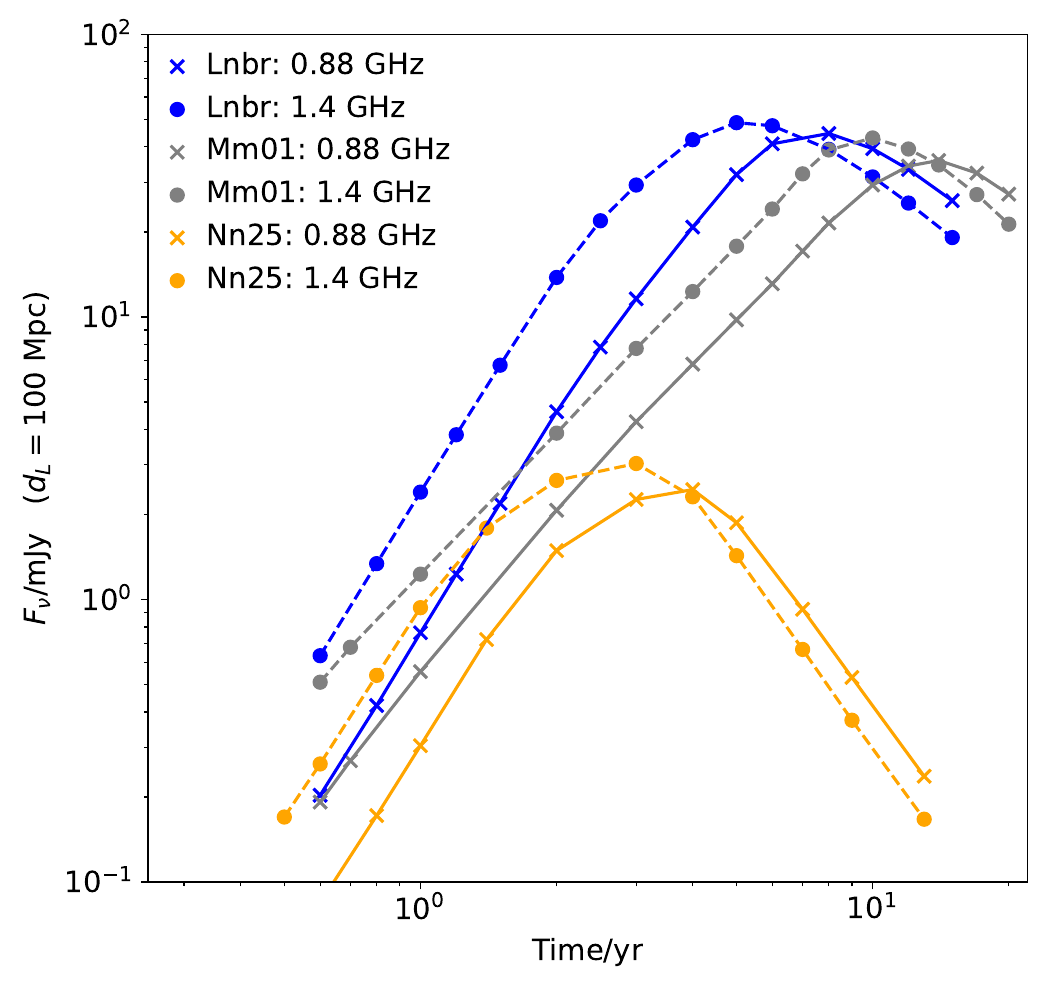}
 \caption{Temporal evolution of the monochromatic flux $F_{\nu}$ in the equatorial direction for model Lnbr, Mn01 and Nn25. The left panel is for $\nu=6$ GHz, and the right panel is for 0.88 and 1.4 GHz.  } 
\label{figA4}
\end{figure*}



\end{document}